\documentclass[aps, prl, reprint,10pt,longbibliography,superscriptaddress]{revtex4-2}

\usepackage[caption=false]{subfig}
\usepackage{amsmath, amssymb, amsfonts, amsthm, bm, mathrsfs, bbm, enumerate}
\usepackage{graphicx}
\usepackage{siunitx}
\usepackage{epstopdf}
\usepackage{color}
\usepackage[usenames,dvipsnames]{xcolor}
\usepackage[citecolor=blue, colorlinks=true, urlcolor=blue, linkcolor=blue]{hyperref}
\usepackage{bbold}
\usepackage{soul}
\usepackage[normalem]{ulem}
\usepackage{floatrow}
\newfloatcommand{capbtabbox}{table}[][\FBwidth]

\newcommand{\eq}[1]{Eq.~(\ref{#1})}

\newcommand{\tab}[1]{Tab.\thinspace{}\ref{#1}}

\newcommand{\fig}[1]{Fig.\thinspace{}\ref{#1}}

\newcommand{\fc}[1]{({#1})}
\newcommand{\figc}[2]{Fig.\thinspace{}\ref{#1}\thinspace{}\fc{#2}}

\newcommand{\exc}[2]{\hat{x}_{\boldsymbol{#1}}^{#2}}

\def \ETH{Institute for Quantum Electronics, ETH Z\"urich, CH-8093 Z\"urich, Switzerland}

\begin{document}

\title{Optical Signatures of Periodic Magnetization: The Moir\'{e} Zeeman Effect}
\author{Alex G\'{o}mez Salvador}
\email{asalvador@phys.ethz.ch}
\affiliation{\ETH}

\author{Clemens Kuhlenkamp}
\affiliation{Department of Physics, Technical University of Munich, 85748 Garching, Germany}
\affiliation{Munich Center for Quantum Science and Technology (MCQST), Schellingstr. 4, D-80799 M{\"u}nchen, Germany}
\affiliation{\ETH}

\author{Livio Ciorciaro}
\affiliation{\ETH}

\author{Michael Knap}
\affiliation{Department of Physics, Technical University of Munich, 85748 Garching, Germany}
\affiliation{Munich Center for Quantum Science and Technology (MCQST), Schellingstr. 4, D-80799 M{\"u}nchen, Germany}
\author{Ata\c{c} \.{I}mamo\u{g}lu}
\affiliation{\ETH}

\begin{abstract}

Detecting magnetic order at the nanoscale is of central interest for the study of quantum magnetism in general, and the emerging field of moir\'{e} magnets in particular. Here, we analyze the exciton band structure that arises from a periodic modulation of the valley Zeeman effect. Despite long-range electron-hole exchange interactions, we find a sizable splitting in the energy of the bright circularly polarized exciton Umklapp resonances, which serves as a direct optical probe of  magnetic order. We first analyze quantum moir\'{e} magnets realized by periodic ordering of electron spins in Mott-Wigner states of transition metal dichalcogenide monolayers or twisted bilayers: we show that spin valley-dependent exciton-electron interactions allow for probing the spin-valley order of electrons and demonstrate that it is possible to observe unique signatures of ferromagnetic order in a triangular lattice and both ferromagnetic and N\'{e}el order in a honeycomb lattice. We then focus on  semiclassical moir\'{e} magnets realized in twisted bilayers of ferromagnetic materials: we propose a detection scheme for moir\'{e} magnetism that is based on interlayer exchange coupling between spins in a moir\'{e} magnet and excitons in a transition metal dichalcogenide monolayer. 
 
\end{abstract}

\date{\today}

\maketitle

Twisted two-dimensional materials have emerged as a fascinating platform to realize exotic phases of matter in solid state systems~\cite{Jarrillo-Herrero2018}. Including twist angles in heterostructures makes it possible to reliably introduce moir\'{e} superlattices, with adjustable lattice constants. This technique has allowed for a number of ground-breaking experiments in magic angle twisted bilayer graphene demonstrating strong electron correlations~\cite{Jarrillo-Herrero2018,Young20,Lu2019,Regan2020,Tang2020,Shimazaki2020,Xu2020,Gao2020}. Recently, twisted bilayers of transition metal dichalcogenides (TMDs) have allowed for the observation of correlated Mott-Wigner states~\cite{Tang2020,Regan2020,Li2021,Shimazaki2021,Wang2020,Zhang2020,Xu2020}. In parallel, it was proposed that twisted two-dimensional magnetic materials would allow for the realization of tunable magnetic phases, called ``moir\'{e} magnets''~\cite{Balents_2020}. While magnetic order in twisted bilayers of magnetic materials has been recently reported~\cite{Xu2021}, their detection poses a significant experimental challenge. To study moir\'{e} magnets it is essential to develop noninvasive probes of magnetism that are sensitive to periodic variations of magnetization on $\sim \SI{10}{\nano\meter}$ length scales.

\smallskip

In this Letter, we propose an all-optical detection scheme that uses exciton resonances as probes for periodic magnetic fields. It has been recently demonstrated that the presence of a periodic potential for excitons with reciprocal lattice vectors $\{ {\textbf G}_i \}$, leads to Bragg-Umklapp scattering of an otherwise dark exciton with momenta $k \in \{ {\textbf G}_i \}$ back to $k=0$, thereby resulting in new bright exciton peaks in the reduced Brillouin zone~\cite{Shimazaki2021,Alexeev2019}. At first glance, it would appear that the presence of large long-range electron-hole exchange interaction~\cite{Glazov2014,WangYao2014,Qiu2015} leading to linearly polarized, finite-momentum excitons would render these Umklapp resonances insensitive to a periodic magnetic potential. Remarkably, we show to the contrary that the excitonic valley Zeeman effect~\cite{Srivastava2015,Xu2015,Heinz2014} ensures large Zeeman splitting of the circularly polarized excitonic Umklapp resonances and thereby provides a unique and direct probe of magnetic order. 

\begin{figure}
    \centering
    \includegraphics[width=8.6cm]{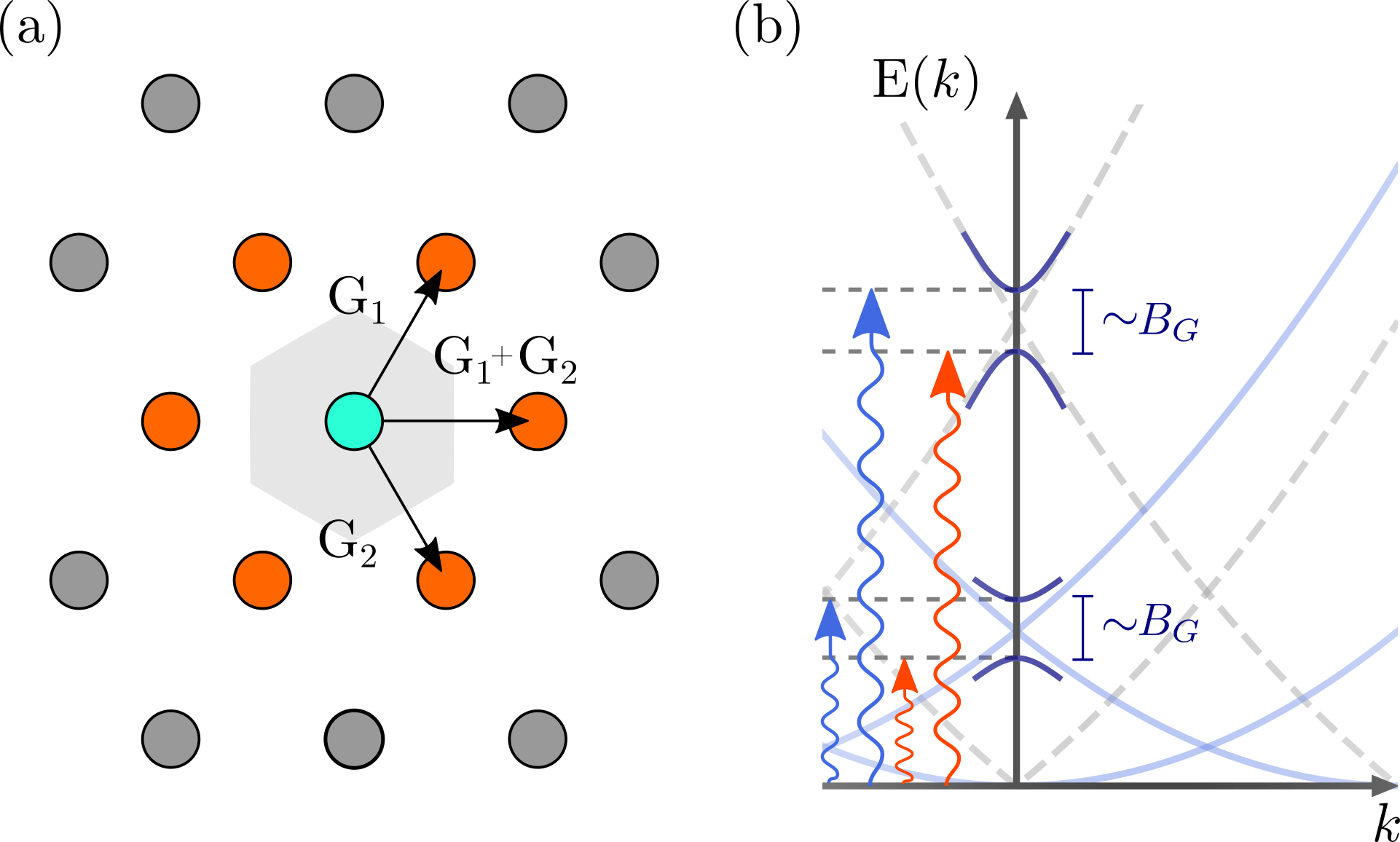}
    \caption{ \fc{a} Momentum space representation of the periodic potential problem, where the area in light gray is the first Brillouin zone. The green dot represents the zero-momentum state, which is connected through the set of reciprocal lattice vectors $\mathcal{G}$ to the first Umklapp states, in orange. \fc{b} Schematic representation of the explored optical resonances. The solid blue (dashed grey) lines correspond to the dispersion of the transversely (linearly) polarized excitonic eigenmode. The red (blue) wavy arrows correspond to resonances belonging to the $\sigma^-$ ($\sigma^+$) subspace.}
    \label{fig:1}
\end{figure}

\begin{figure*}
   \centering
   \includegraphics[width=17.2cm]{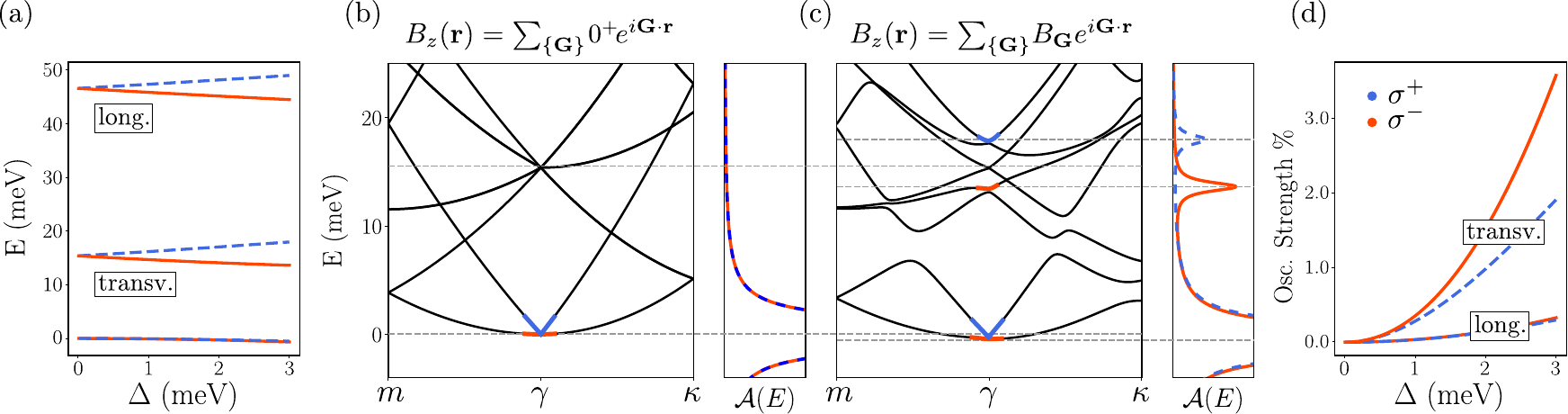}
   \caption{ \fc{a} Energy spectrum of the two circularly polarized subspaces as a function of the periodic magnetic field amplitude $\Delta=g\mu_B |B_\textbf{G}|$, assuming vanishing average field. The energy of the states belonging to the $\sigma^-$ ($\sigma^+$) subspace is shown in solid red (dashed blue) lines. As $\Delta$ increases, the Umklapp states notably split, while the zero-momentum excitons remain quasidegenerate. \fc{b,c} Exciton band structure for a periodic magnetic field with vanishingly small \fc{b} and finite strength yielding $\Delta=\SI{3}{meV}$ \fc{c}. We denote the high-symmetry points of the superlattice with $m$, $\gamma$, and $\kappa$. The degeneracy of the lowest energy Umklapp scattered bands in \fc{b} is lifted for $\Delta \neq 0$ in \fc{c}. To the right, the corresponding zero-momentum spectral functions for $\sigma^+$ and $\sigma^-$ subspaces in dashed blue and solid red respectively. For $\Delta=0$, the spectral functions present a single degenerate peak for each zero-momentum exciton. For $\Delta\neq0$, the transverse Umklapp resonances in each subspace become bright, resulting in additional peaks in $\mathcal{A}(E)$. The remarkable splitting of the Umklapp peaks is a unique signature of the moir\'{e} Zeeman effect. \fc{d} Oscillator strengths of the finite quasimomentum states as a function of $\Delta$, with $a = \SI{10}{nm}$. In solid red lines (dashed blue lines), the results corresponding to the $\sigma^-$ ($\sigma^+$) subspace. For $\Delta\gtrsim\SI{1.5}{\milli\electronvolt}$ the two resonances are bright enough to act as a direct optical probe of the periodic magnetic field.}
   \label{fig: 2}
\end{figure*}


\textbf{Excitons in periodic magnetic potentials.---}
\vspace{0.1cm}

Let $\exc{k}{+}$ ($\exc{k}{-}$) be the bosonic annihilation operators of bright excitons with momentum $\boldsymbol{k}=(k_x,k_y)$ in the K (K') valley. Henceforth, we refer to the valley degree of freedom as the exciton pseudospin, where K (K') excitons have pseudospin $+$ $(-)$. In the presence of long-range electron-hole exchange interactions, the effective Hamiltonian for excitons with momentum $\boldsymbol{k}$ in the vicinity of the K and K' valleys reads
\small
\begin{equation}
    \hat{\mathcal{H}}_{\boldsymbol{k}}=\begin{pmatrix}\hat{x}^+_{\boldsymbol{k}}\\\hat{x}^{-}_{\boldsymbol{k}}\end{pmatrix}^\dagger\bigg{[}\frac{\hbar^2k^2}{2m_X} \mathbb{1} +\frac{k\mathcal{J}}{|\text{K}|}\begin{pmatrix}1 & e^{-i2\theta_{\boldsymbol{k}}}\\e^{i2\theta_{\boldsymbol{k}}}&1\end{pmatrix}\bigg{]}\begin{pmatrix}\hat{x}^+_{\boldsymbol{k}}\\\hat{x}^{-}_{\boldsymbol{k}}\end{pmatrix},
    \label{eq:1}
\end{equation}
\normalsize
where $\tan\theta_{\boldsymbol{k}}=k_y/k_x$, $m_X$ is the exciton mass, $\mathcal{\mathcal{J}}$ characterizes the strength of the electron-hole exchange, and $|\text{K}|=4\pi/3a_0$ is the TMD valley momentum, with $a_0\simeq 3$\r{A} the corresponding lattice constant~\cite{WangYao2014,WangYao_Review}. The Hamiltonian in \eq{eq:1} can be diagonalized through  $\hat{\mathcal{H}}_{\boldsymbol{k}}=U_{\boldsymbol{k}}\: \text{diag}[\lambda_{\boldsymbol{k},l},\lambda_{\boldsymbol{k},t}]U_{\boldsymbol{k}}^\dagger$, where $\lambda_{\boldsymbol{k},(l,t)}=\frac{\hbar^2k^2}{2m_x}+\frac{k\mathcal{J}}{|\text{K}|}\pm \frac{k\mathcal{J}}{|\text{K}|}$ and $U_{\boldsymbol{k}}=\frac{1}{\sqrt{2}}\begin{pmatrix} e^{-i\theta_{\boldsymbol{k}}} & e^{i\theta_{\boldsymbol{k}}} \\ e^{-i\theta_{\boldsymbol{k}}} & -e^{i\theta_{\boldsymbol{k}}}\end{pmatrix}.$ The eigenmodes of the system are equal weight superpositions of excitons in both valleys and have longitudinal ($l$) and transverse ($t$) polarizations, with annihilation operators $\hat{x}^{l,t}_{\boldsymbol{k}}$~\cite{Qiu2015,Glazov2014,Wang2020}. As a consequence, the pseudospin of the eigenmodes lie in plane. 

\smallskip

We consider a triangular magnetic lattice, with an effective field perpendicular to the TMD plane and with vanishing average, $B_z(\boldsymbol{r})=\sum_{\{\textbf{G}\}} B_{\textbf{G}}e^{i\textbf{G}\cdot\boldsymbol{r}}$. We work in the regime where the lattice constant $a$ of the magnetic potential is much larger than $a_0$, such that the low energy continuum model of \eq{eq:1} applies. The periodic magnetic field then interacts with excitons through an effective valley Zeeman coupling
\begin{equation}
    \hat{\mathcal{V}}_{\boldsymbol{k}}=\frac{g\mu_B}{2}\sum_{\textbf{G}}B_{\textbf{G}} \begin{pmatrix}  \hat{x}^{+}_{\boldsymbol{k}+\textbf{G}}\\ \hat{x}^{-}_{\boldsymbol{k}+\textbf{G}} \end{pmatrix} ^\dagger \begin{pmatrix} 1 & 0 \\ 0 & -1 \end{pmatrix} \begin{pmatrix} \hat{x}^{+}_{\boldsymbol{k}} \\ \hat{x}^{-}_{\boldsymbol{k}}\end{pmatrix} + \text{H.c.}\;,
    \label{eq:2}
\end{equation}
where $g$ is the exciton $g$ factor ($g \simeq 4.3$ for MoSe$_2$ excitons). Consequently, the full Hamiltonian of the system is $\hat{\mathcal{H}}=\sum_{\boldsymbol{k}}\hat{\mathcal{H}}_{\boldsymbol{k}}+\hat{\mathcal{V}}_{\boldsymbol{k}}$. As described earlier, we are interested in the polarization splitting of Bragg-scattered excitons, which become optically active by mixing with zero-momentum ($k=0$) excitons. Experimentally relevant magnetic fields are often small, $\mu_BB_{\textbf{G}}\ll\lambda_{\textbf{G},l},\lambda_{\textbf{G},t}$, such that a nearly free exciton model applies. The relevant contributions of $B_z(\boldsymbol{r})$ stem from the first Umklapp states in \fig{fig:1}, labeled by $\mathcal{G}=\{ \pm\textbf{G}_1,\pm\textbf{G}_2, \pm(\textbf{G}_1+\textbf{G}_2)\}$, where  $\textbf{G}_1=\frac{2\pi}{\sqrt{3}a}(1,\sqrt{3})$ and $\textbf{G}_2=\frac{2\pi}{\sqrt{3}a}(1,-\sqrt{3})$. Furthermore, in this limit, it is justified to consider a truncated Hilbert space containing the $k=0$ state and the six first Umklapp states, each with two polarizations. 

\smallskip

We focus on a featureless triangular lattice, which has $C_6$ symmetry and the Hamiltonian commutes with $\pi/3$ rotations generated by the $z$ component of the total angular momentum $\hat{J}_z=\hat{L}_z\otimes\mathbb{1}_{2\times2}+\mathbb{1}\otimes 2\hat{S}_z$. Here $\hat{L}_z$ denotes orbital angular momentum and $\hat{S}_z$ the valley pseudospin. Consequently, $\sigma^+$ ($J_z = 1$) and  $\sigma^-$ ($J_z = -1$) subspaces decouple, each containing states that transform identically to $\hat{x}^{+}_{0}$ and $\hat{x}^{-}_{0}$, respectively. These two subspaces are spanned by $\sigma^+=\{ \hat{x}^{+}_0, \hat{x}^+_{\text{G},l},\hat{x}^+_{\text{G},t} \}$ and $\sigma^-=\{ \hat{x}^{-}_0,\hat{x}^-_{\text{G},l}, \hat{x}^-_{\text{G},t} \}$, with
\begin{equation}
\begin{split}
    \hat{x}^-_{\text{G},l}=\frac{1}{\sqrt{6}}\sum_{\textbf{G}}\hat{x}^l_{\textbf{G}}e^{-i\theta_{\textbf{G}}}, \quad \hat{x}^+_{\text{G},l}=\frac{1}{\sqrt{6}}\sum_{\textbf{G}}\hat{x}^l_{\textbf{G}}e^{i\theta_{\textbf{G}}},\\
    \hat{x}^-_{\text{G},t}=\frac{1}{\sqrt{6}}\sum_{\textbf{G}}\hat{x}^t_{\textbf{G}}e^{-i\theta_{\textbf{G}}}, \quad \hat{x}^+_{\text{G},t}=\frac{1}{\sqrt{6}}\sum_{\textbf{G}}\hat{x}^t_{\textbf{G}}e^{i\theta_{\textbf{G}}}.
    \label{eq:3}
\end{split}
\end{equation}
In this basis, the Hamiltonian is block diagonal and takes the form
\footnotesize
\begin{equation}
    \begin{split}
    \hat{\mathcal{H}}=&\begin{pmatrix} \hat{x}^+_{0} \\ \hat{x}^+_{\text{G},l} \\ \hat{x}^+_{\text{G},t} \end{pmatrix}^{\dagger} \begin{pmatrix} 0 & \frac{\sqrt{3}\Delta}{2} & \frac{\sqrt{3}\Delta}{2}\\ \frac{\sqrt{3}\Delta}{2} & \lambda_{\textbf{G},l}+\frac{3\Delta}{4} & \frac{\Delta}{4} \\ \frac{\sqrt{3}\Delta}{2} & \frac{\Delta}{4}& \lambda_{\textbf{G},t}+\frac{3\Delta}{4} \end{pmatrix}\begin{pmatrix} \hat{x}^+_{0} \\ \hat{x}^+_{\text{G},l} \\ \hat{x}^+_{\text{G},t} \end{pmatrix} + \\
    +&\begin{pmatrix} \hat{x}^-_{0} \\ \hat{x}^-_{\text{G},l} \\ \hat{x}^-_{\text{G},t} \end{pmatrix}^{\dagger} \begin{pmatrix} 0 & -\frac{\sqrt{3}\Delta}{2} & \frac{\sqrt{3}\Delta}{2} \\ -\frac{\sqrt{3}\Delta}{2} & \lambda_{\textbf{G},l}-\frac{3\Delta}{4} & \frac{\Delta}{4} \\ \frac{\sqrt{3}\Delta}{2} & \frac{\Delta}{4} & \lambda_{\textbf{G},t}-\frac{3\Delta}{4} \end{pmatrix} \begin{pmatrix} \hat{x}^-_{0} \\ \hat{x}^-_{\text{G},l} \\ \hat{x}^-_{\text{G},t} \end{pmatrix},
    \label{eq:4}
    \end{split}
\end{equation}
\normalsize
with $\Delta=g\mu_B|B_\textbf{G}|$ and $|B_\textbf{G}|$ the amplitude of the first Fourier component of the periodic magnetic field. We set $\mathcal{J} = \SI{300}{meV}$ and $m_X = 1.3\,m_e$, motivated by the values estimated for MoSe$_2$ monolayers, and fix the lattice constant of the periodic magnetic field to $a=\SI{10}{nm}$~\cite{Smolenski2021}. We then find the eigenstates of \eq{eq:4} as a function of $\Delta$, which are presented in \figc{fig: 2}{a}. Remarkably, we show that the energies of the Umklapp states split linearly with $\Delta$, while the zero-momentum excitons remain quasi-degenerate. Intuitively, the main $k=0$ resonances are homogeneous in real space and to first order only feel the average magnetic field, which we assumed to be zero. 
\smallskip

\figc{fig: 2}{b} and \figc{fig: 2}{c} show the excitonic band structure with an infinitesimally small and a finite periodic magnetic field, respectively. Next to the band structure, we depict the zero-momentum spectral functions $\mathcal{A}(E)$ for the $\sigma^-$ and $\sigma^+$ subspaces which are directly accessible via optical spectroscopy. In contrast to nonmagnetic potentials, the Umklapp resonances at the $\gamma$ point exhibit a pronounced splitting~\cite{Smolenski2021,Shimazaki2021}. The emergence of additional peaks in the exciton spectrum serves as a direct signature of periodic magnetic structures, while their splitting distinguishes them from the combination of a purely electronic periodic potential and a uniform magnetic field. The splitting can be intuitively understood as arising from nonzero coupling between the superposed Umklapp states of each symmetry basis state. Consequently, it is explicitly dependent on the lattice geometry and determined by $\frac{\Delta}{2}(1+\cos{2\delta\theta_\textbf{G}})$, with $\delta\theta_\textbf{G}$ the angle between consecutive momentum vectors that form the symmetry basis element, depicted in \figc{fig:1}{a}; for the particular scenario of the triangular lattice we have $\delta\theta_\textbf{G} = \frac{\pi}{3}$.

\smallskip

We compute the oscillator strengths of the Umklapp resonances as a function of the potential strength in \figc{fig: 2}{d}. We find that the brightest Umklapp peaks, which originate from Bragg scattering of transversely polarized excitons, have a relative oscillator strength of $\simeq \SI{1}{\percent}$ for $\Delta = \SI{1.5}{meV}$ ($|B_\textbf{G}|\simeq\SI{12}{T}$). Provided that the exciton linewidth is predominantly due to radiative decay, this splitting can be easily observed~\cite{Shimazaki2021,Smolenski2021}. This scheme is suitable to detect periodic modulations with small moir\'{e} lattice constants of 5--\SI{20}{nm}, which are inaccessible to established imagining techniques such as nitrogen-vacancy magnetometry~\cite{Maletinsky2012}\footnote{Since the strength of the electron-hole exchange interaction depends on the dielectric constant, its precise value will strongly vary for each system. For $\mathcal{J}\gtrsim \SI{250}{meV}$, $\lambda_{\boldsymbol{G},l}\gg \lambda_{\boldsymbol{G},t}$ and the longitudinal modes couple weakly to the zero-momentum states due to the large energy detuning. Consequently, there are only two bright Umklapp exciton resonances corresponding to the transverse polarized exciton branch. For $\mathcal{J} \lesssim 50$~meV, $\lambda_{\boldsymbol{G},l} \sim \lambda_{\boldsymbol{G},t}$ and the longitudinal modes also become bright, resulting in four Umklapp resonances.}.

\begin{figure*}
    \centering
    \includegraphics[width=17.2cm]{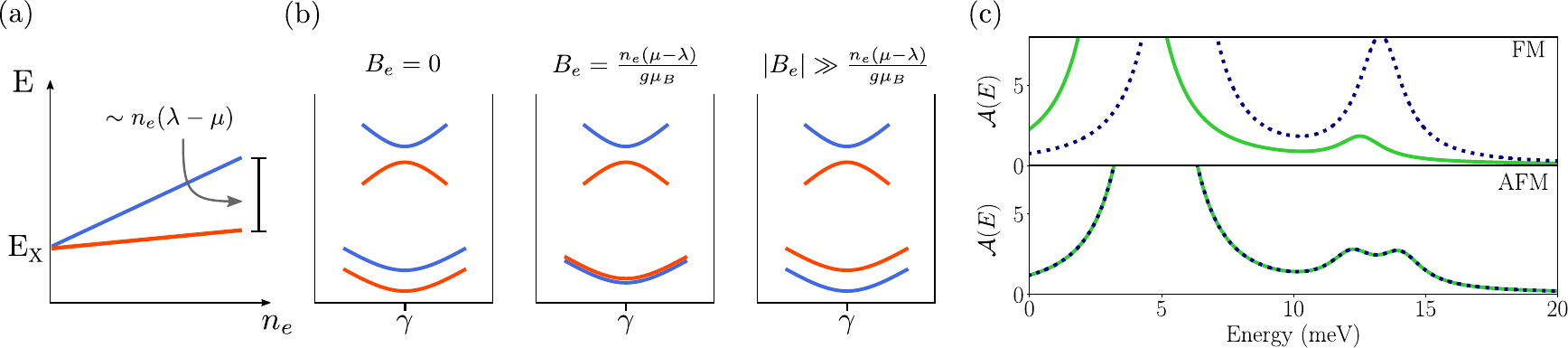}
    \caption{\fc{a} Schematic of the exciton energy shift as a function of electron density. The difference between the $\sigma^+$ and $\sigma^-$ exciton is proportional to the effective constant magnetic field due to valley Zeeman effect. \fc{b} Schematic of the exciton band structures around the $\gamma$ point of the ferromagnetic triangular lattice for different values of constant external magnetic field $B_e$ and $\lambda\neq\mu$. The robustness of the Umklapp peaks against constant magnetic fields allows for direct manipulation of the zero-momentum resonances. \fc{c} Spectral functions for the ferromagnetic (FM) and antiferromagnetic (AFM) honeycomb lattice, where the solid green (dotted navy blue) lines correspond to the $\sigma^-$ $(\sigma^+)$ subspace. We consider a moir\'{e} lattice constant of $a=\SI{15}{\nano\meter}$ and the experimentally relevant parameters $n_e\lambda=\SI{2}{\milli\electronvolt}$ and $n_e\mu=\SI{4}{meV}$. The net magnetization in the FM leads to a large splitting of the main exciton resonances, while $\sigma^+$ and $\sigma^-$ excitons remain degenerate in the AFM. Moreover, although both scenarios present energetically split Umklapp peaks, the signatures are notably distinct. The AFM presents two Umklapp peaks for each subspace, which are subspace degenerate. On the other hand, the FM has single Umklapp peaks, which are not subspace degenerate.}
    \label{fig: 3}
\end{figure*}


\vspace{0.75cm}
\textbf{Umklapp signatures of a ferromagnetic triangular lattice.---}
\vspace{0.1cm}

Recent experiments have demonstrated the possibility of generating periodic charge distributions at low carrier density~\cite{Feng-Wang2020,Mak-Shan2020,Shimazaki2021,Smolenski2021,Zhou21}. The emergence of excitonic Umklapp resonances in optical spectroscopy due to exciton-electron interactions has served as a direct probe of broken translational invariance of electrons~\cite{Smolenski2021,Shimazaki2021}. In TMDs, excitons and electrons interact differently depending on whether they are in the same or opposite valley~\cite{Fey_2020}. Moreover, large spin-orbit coupling ensures that for low electron density, the spin and valley indices are locked~\cite{WangYao_Review}. Consequently, it should be possible to detect spin order in triangular Mott-Wigner lattices.

\smallskip

The interaction between excitons and electrons in different valleys leads to the formation of attractive and repulsive exciton-polarons: we refer to the effective (intervalley) interaction strength for the repulsive-polaron branch as $\mu$. For excitons occupying the same valley as the electrons, both phase space filling and repulsive interactions contribute to the electron density-dependent energy shift with strength $\lambda$. Henceforth, we consider moir\'{e} lattice constants of 5-\SI{25}{\nano\meter}, larger than the trion Bohr radius $\sim$\SI{2}{\nano\meter}. In this regime, resonantly excited excitons (repulsive polarons) interact repulsively and weakly with electrons and a Hartree approximation with a contact interaction provides a good description \cite{Fey_2020}. We therefore write the exciton-electron interaction as
\begin{align}
    \hat{\mathcal{H}}_{\text{int}}= \int d^2 r\;  \bigg{\{}&\left[\lambda n_e^+(\bm r) + \mu n_e^-(\bm r)\right] \hat{x}_+^\dagger(\textbf{r}) \hat{x}_+(\textbf{r})\notag \\
    +&\left[\mu n_e^+(\bm r) + \lambda n_e^-(\bm r)\right] \hat{x}_-^\dagger(\textbf{r}) \hat{x}_-(\textbf{r})\bigg{\}},
\end{align}
where $n_e^{\pm}(\bm r)$ is the electron density in the K $(+)$ and K' $(-)$ valley. For simplicity, we consider a ferromagnetic (FM) triangular lattice of electrons, which are polarized in the K valley. When $n_e^-(\bm r)=0$, the Hamiltonian can be recast into
\begin{align}
    \nonumber &\hat{\mathcal{H}}_{\text{int}}^{\text{FM}}=\int d^2r \; n_e(\textbf{r}) \; \times \\ \begin{pmatrix} \hat{x}_+(\textbf{r}) \\ 
    \hat{x}_-(\textbf{r}) \end{pmatrix}^\dagger & \bigg{(}\frac{\lambda+\mu}{2}\mathbb{1}+\frac{\lambda-\mu}{2}\sigma_z\bigg{)\begin{pmatrix} \hat{x}_+(\textbf{r}) \\ \hat{x}_-(\textbf{r}) \end{pmatrix}},
    \label{eq: 5}
\end{align}
where $\mathbb{1}$ and $\sigma_z$ are the 2$\times$2 identity and the $z$-Pauli matrix respectively. The interaction Hamiltonian is now a sum of a periodic identity potential that couples equally strongly to both exciton valley pseudospins and an effective periodic magnetic potential. If $\lambda = \mu$, we recover the limit where an exciton couples to the periodic charge distribution in a pseudospin-independent way and both the $k=0$ as well as the Umklapp resonances remain degenerate but experience a polaron blue shift. If on the other hand $\mu > \lambda$, as the experiments in monolayer  MoSe$_2$ demonstrate, excitons feel an effective spatially modulated field both proportional to $\mu - \lambda$~\cite{Back17}, as depicted in \figc{fig: 3}{a,b}. Therefore, in the presence of periodic ordering of electrons in a single valley, both the $k=0$ and the Umklapp resonances present splittings on the order of $n_e|\mu - \lambda|$, where $n_e$ is the average electron density. 

\smallskip

The ferromagnetic Mott-Winger state in a triangular lattice therefore differs from the generic scenario we analyzed since the effective periodic magnetic field has a nonzero average $\langle B({\bf r}) \rangle \neq 0$. If we apply an external homogeneous field $B_e$ with strength $\mu_BgB_e/2=n_e(\lambda - \mu)/2$, we recover the case of a periodic field with zero spatial average: in this case, only the Umklapp resonances, which are insensitive to $B_e$, retain a finite splitting. In the limit $|\mu_BgB_e/2|\gg n_e(\mu-\lambda)/2$ we obtain an anomalous regime where the closest zero-momentum and Umklapp states belong to the same subspace (\figc{fig: 3}{b}). State-of-the-art experiments should allow for detection of  Umklapp resonances with oscillator strengths $\gtrsim$ 1\% relative to the main excitonic peak~\cite{Smolenski2021}. Furthermore, the splitting of the Umklapp states can be resolved for $n_e(\mu - \lambda)\simeq$ \SI{1}{\milli\electronvolt}. We expect both conditions to be readily satisfied for moir\'{e} lattice constants $\le 20$~nm and typical exciton-electron interactions strengths.


\smallskip

\textbf{Umklapp signatures of ferromagnetic and antiferromagnetic honeycomb lattices.---}
\vspace{0.11cm}

While electronic triangular lattices occur naturally in Wigner crystals, electrons can also spontaneously organize in honeycomb lattices to form mutually stabilized Mott-Wigner states in twisted homobilayer TMDs. Although super-exchange interactions are expected to generate antiferromagnetic order in bipartite lattices, competition with direct exchange can favor ferromagnetism~\cite{Morales-Duran21}.

To identify the  actual spin order using Umklapp signatures we take
into account the $C_3$ symmetry of honeycomb lattices by developing an analogous model to the one previously discussed, considering two enlarged five-dimensional subspaces; see the Supplemental Material~\cite{Supplement} for details. The ferromagnetic (FM) and antiferromagnetic (AFM) scenarios can be recast in expressions akin to Eq.~(\ref{eq: 5}). For both cases we identify an identity and a magnetic potential, which are proportional to $\lambda+\mu$ and $\lambda-\mu$ respectively. 

\smallskip

For the FM case, analogously to the triangular lattice, we find two transverse Umklapp states despite the reduced symmetry of the lattice. This can be understood from the potential excitons feel, which is $C_6$ symmetric and allows the states in each subspace to hybridize into a bright and a dark Umklapp state. As before, we observe a splitting of the Umklapp peaks that is linearly proportional to the effective magnetic field strength, see \figc{fig: 3}{c} where we plot the spectral function for experimentally relevant parameters $n_e\lambda=\SI{2}{\milli\electronvolt}$ and $n_e\lambda=\SI{4}{\milli\electronvolt}$; see Supplemental Materials for details~\cite{Supplement}. In this scenario, the splitting of the main resonances is due to a combination of the nonzero average magnetic field and the hybridization with the Umklapp states. While the difference in interaction strengths $\lambda-\mu$, which determines the Umklapp peak splitting, is sizable in experiments, it may be further increased using Feshbach resonances in multilayer structures that selectively enhance interactions between excitons and electrons in different valleys~\cite{Shimazaki2021,Kuhlenkamp21}.

\smallskip

In the AFM case, the identity and magnetic potentials are proportional to $(\lambda+\mu)(n_e^+(\textbf{r})+n_e^-(\textbf{r}))$ and $(\lambda-\mu)(n_e^+(\textbf{r})-n_e^-(\textbf{r}))$ respectively. The fact that the two potentials present a different spatial profile results, in general, in  $C_3$-symmetric potentials and in two bright transverse Umklapp peaks for each subspace, see \figc{fig: 3}{c}. For vanishing average magnetic fields, the subspaces are degenerate even in the presence of an effective periodic magnetic field \footnote{In the presence of an external constant magnetic field, the main resonances shift due to the Zeeman effect while the Umklapp peaks remain almost degenerate}. The striking differences in the excitonic spectral functions of the FM and AFM scenarios allow for determining the nature of spin-valley order using far-field optical measurements.

\begin{figure}
    \centering
    \includegraphics[width=8.6cm]{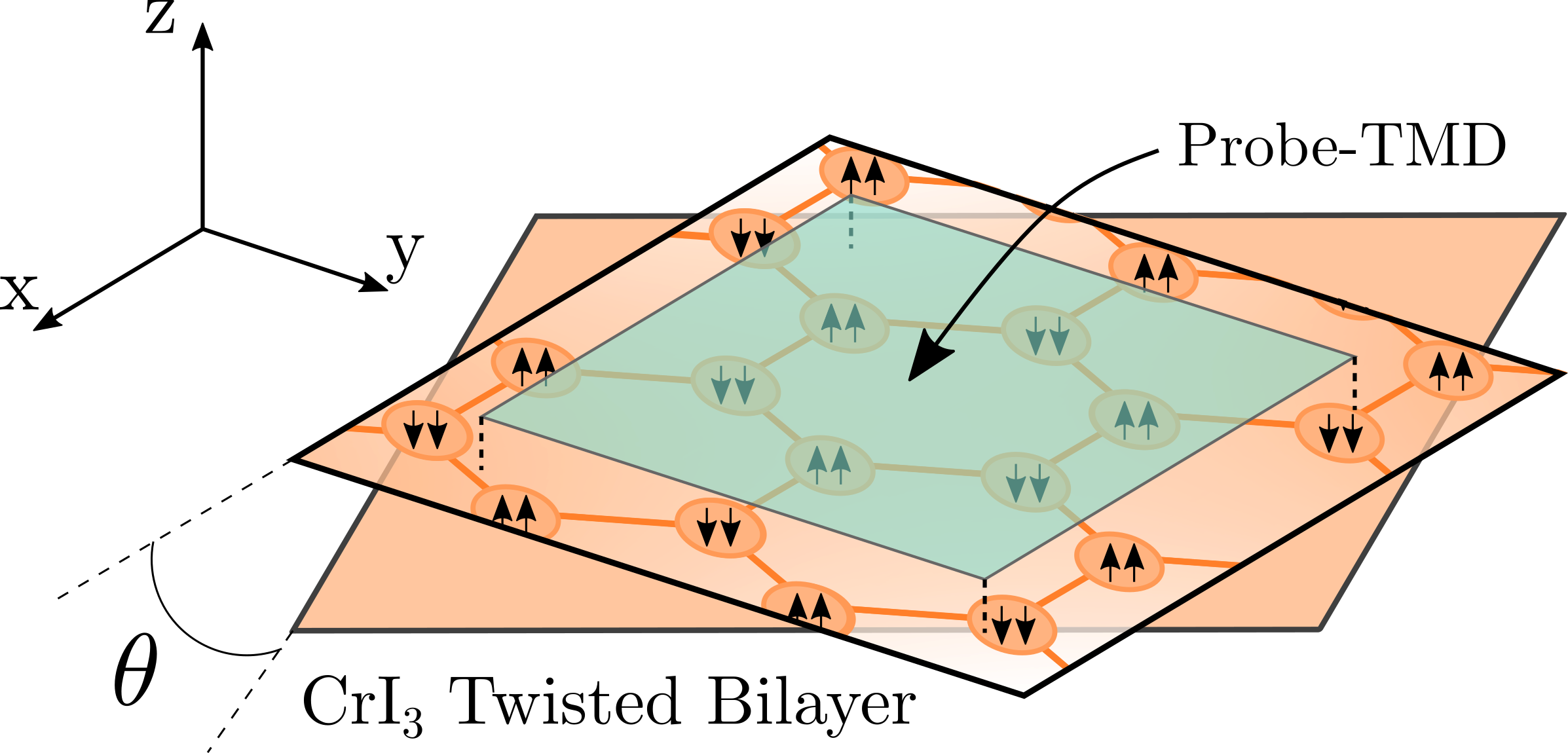}
    \caption{Schematic representation of a CrI$_3$ twisted bilayer with a generic spin distribution in the top layer. On top of the bilayer, the probe-TMD monolayer in green.}
    \label{fig: 4}
\end{figure}


\smallskip

\textbf{Twisted bilayers of CrI$_3$ .---}
\vspace{0.1cm}

The methods developed here can also be used to detect semiclassical magnetic order, which has recently been predicted to appear in twisted bilayers of certain TMD materials due to stacking dependent interlayer exchange couplings~\cite{Balents_2020}. We focus on the special case of CrI$_3$ bilayers and show that a MoSe$_2$ monolayer can be used to sense the magnetic textures, by virtue of an effective valley Zeeman effect~\cite{Zhong2020,Livio2020}, see~\fig{fig: 4} for an illustration. We find that for twist angles $\theta\lesssim\SI{5}{\degree}$, ordered magnetic patterns are detectable by exciton scattering. The Supplemental Material~\cite{Supplement} provides detailed calculations.

For experimentally accessible values of $\theta=\SI{1.5}{\degree}$ and a Zeeman coupling $\delta=\SI{3}{meV}$, our analysis shows that two bright Umklapp resonances with a relative splitting of $\sim \SI{3}{\milli\electronvolt}$ and relative oscillator strengths of around \SI{1.6}{\percent} and \SI{2.2}{\percent} emerge. Although one of the resonances falls within the linewidth of the main resonance for our chosen parameters, the second one presents an energy difference large enough to be spectrally resolved.


\smallskip

\textbf{Conclusions and outlook.---}
\vspace{0.1cm}

We have studied the optical response of TMD excitons to periodic magnetic fields generated either due to spin order in Mott-Wigner states in the same monolayer or due to semiclassical moir\'{e} magnets in a neighboring magnetic bilayer, using a weak coupling model. While nonmagnetic potentials give rise to degenerate Umklapp peaks for both $\sigma^+$ and $\sigma^-$ polarized excitons, periodic magnetic potentials split the Umklapp resonances and provide a clear signature of a magnetic lattice. We have shown how these signatures can be used as a direct probe for quantum magnetism in triangular and honeycomb moir\'{e} lattices and for moir\'{e} magnets arising in twisted bilayers of CrI$_3$.

We thank T. Smole\'{n}ski, I. Schwartz, Y. Shimazaki, A. M\"{u}ller and P. Dolgirev for insightful discussions.


%


\appendix
\newpage
\setcounter{figure}{0}
\setcounter{equation}{0}

\renewcommand{\thepage}{S\arabic{page}} 
\renewcommand{\thesection}{S\arabic{section}} 
\renewcommand{\thetable}{S\arabic{table}}  
\renewcommand{\thefigure}{S\arabic{figure}} 
\renewcommand{\theequation}{S\arabic{equation}} 

\onecolumngrid

\begin{center}
\textbf{\Large{\large{Supplemental Material: \\ Optical signatures of periodic magnetization: the moiré Zeeman effect}}}
\end{center}

\title{Supplemental Material: Optical signatures of periodic magnetization: the moiré Zeeman effect}


\section{Ferromagnetic triangular lattice}

To study the scattering of an exciton off a polarised triangular electron lattice, we have reduced the problem to the sum of a periodic identity potential and a periodic magnetic potential in \eq{eq: 5} of the main text. While the weak-coupling model for the magnetic potential has been carried out in the first section of the main text, we have to develop a similar formalism for the periodic identity potential. Here we present the general weak coupling model and apply it to a ferromagnetic triangular lattice.

\indent

The excitonic eigenmodes in the presence of long-range electron-hole exchange present longitudinal and transverse polarisations \begin{equation}
    \hat{\mathcal{H}}_0=\sum_{\boldsymbol{k}}\begin{pmatrix}\hat{x}_{\boldsymbol{k},l}\\\hat{x}_{\boldsymbol{k},t}\end{pmatrix}^\dagger\begin{pmatrix}\lambda_{\boldsymbol{k},l} & 0 \\ 0 & \lambda_{\boldsymbol{k},t} \end{pmatrix}\begin{pmatrix}\hat{x}_{\boldsymbol{k},l}\\\hat{x}_{\boldsymbol{k},t}\end{pmatrix}
\end{equation}
and we consider that excitons interact with a $C_6$-symmetrical periodic potential with interaction strength $\Delta$ of the form
\begin{equation}
    \hat{\mathcal{V}}=\Delta\sum_i\cos(\textbf{G}_i\textbf{r})\begin{pmatrix}  \hat{x}_{+}(\textbf{r})\\ \hat{x}_{-}(\textbf{r}) \end{pmatrix} ^\dagger \begin{pmatrix}1&0\\0&\pm1\end{pmatrix} \begin{pmatrix}  \hat{x}_{+}(\textbf{r})\\ \hat{x}_{-}(\textbf{r}) \end{pmatrix}.
\end{equation}
We refer to the potential with the $+$ ($-$) sign as an identity (magnetic) potential. In the weak-coupling limit, the relevant contributions come from the first Umklapp states $\mathcal{G}$, justifying the truncation of the Hilbert space to zero momentum and first Umklapp states. Then, the interaction in momentum space reads
\begin{equation}
    \hat{\mathcal{V}}=\frac{\Delta}{2}\sum_{\substack{\boldsymbol{k}\\\textbf{G}\in\mathcal{G}}} \begin{pmatrix} \hat{x}_{\boldsymbol{k}+\textbf{G},+}\\ \hat{x}_{\boldsymbol{k}+\textbf{G},-} \end{pmatrix} ^\dagger \begin{pmatrix} 1 & 0 \\ 0 & \pm 1 \end{pmatrix} \begin{pmatrix}\hat{x}_{\boldsymbol{k},+} \\ \hat{x}_{\boldsymbol{k},-} \end{pmatrix} + \text{h.c.} \:.
    \label{eq:S3}
\end{equation}
Following the arguments presented in the main text, the Hamiltonian $\hat{\mathcal{H}}=\hat{\mathcal{H}}_0+\hat{\mathcal{V}}$ is block diagonal in the symmetry basis and takes the form
\begin{equation}
    \begin{split}
    \hat{\mathcal{H}}=&\begin{pmatrix} \hat{x}_{0,+} \\ \hat{x}^+_{\text{G},l} \\ \hat{x}^+_{\text{G},t} \end{pmatrix}^{\dagger} \begin{pmatrix} 0 & \frac{\sqrt{3}}{2}\Delta & \frac{\sqrt{3}}{2}\Delta \\ \frac{\sqrt{3}}{2}\Delta & \lambda_{\textbf{G},l}+\frac{2\mp1}{4}\Delta & \frac{2\pm1}{4}\Delta \\ \frac{\sqrt{3}}{2}\Delta & \frac{2\pm1}{4}\Delta & \lambda_{\textbf{G},t}+\frac{2\mp1}{4}\Delta \end{pmatrix}\begin{pmatrix} \hat{x}_{0,+} \\ \hat{x}^+_{\text{G},l} \\ \hat{x}^+_{\text{G},t} \end{pmatrix} + \\
    +&\begin{pmatrix} \hat{x}_{0,-} \\ \hat{x}^-_{\text{G},l} \\ \hat{x}^-_{\text{G},t} \end{pmatrix}^{\dagger} \begin{pmatrix} 0 & \pm\frac{\sqrt{3}}{2}\Delta & \mp\frac{\sqrt{3}}{2}\Delta \\ \pm\frac{\sqrt{3}}{2}\Delta & \lambda_{\textbf{G},l}+\frac{-1\pm2}{4}\Delta & \frac{-1\mp2}{4}\Delta \\ \mp\frac{\sqrt{3}}{2}\Delta & \frac{-1\mp2}{4}\Delta & \lambda_{\textbf{G},t}+\frac{-1\pm2}{4}\Delta \end{pmatrix} \begin{pmatrix} \hat{x}_{0,-} \\ \hat{x}^-_{\text{G},l} \\ \hat{x}^-_{\text{G},t} \end{pmatrix}.
    \end{split}
    \label{Seq:4}
\end{equation}
Note that when keeping the lower sign we recover \eq{eq:4} of the main text. We are also interested in taking into account not only purely periodic terms but also constant contributions. For a constant magnetic field the interaction in the excitonic eigenbasis results in
\begin{equation}
    \begin{pmatrix}\hat{x}_{\boldsymbol{k},+} \\ \hat{x}_{\boldsymbol{k},-} \end{pmatrix}^\dagger\begin{pmatrix}1&0\\0&-1\end{pmatrix} \begin{pmatrix}\hat{x}_{\boldsymbol{k},+} \\ \hat{x}_{\boldsymbol{k},-} \end{pmatrix}=\begin{pmatrix}\hat{x}_{\boldsymbol{k},l}\\\hat{x}_{\boldsymbol{k},t}\end{pmatrix}^\dagger \begin{pmatrix}0&1\\1&0\end{pmatrix}\begin{pmatrix} \hat{x}_{\boldsymbol{k},l}\\\hat{x}_{\boldsymbol{k},t}\end{pmatrix}.
    \label{Seq:5}
\end{equation}

Taking into account the electron density profile to be $n_e(\textbf{r})=\frac{2n_e}{3}\big{\{}\cos{\big{[}\text{G}\big{(}\frac{x}{2}+\frac{\sqrt{3}y}{2}\big{)}\big{]}}+\cos{\big{[}\text{G}\big{(}\frac{x}{2}-\frac{\sqrt{3}y}{2}\big{)}\big{]}}+\cos{(\text{G}x)}+\frac{3}{2} \big{\}}$, where G is the modulus of the three reciprocal lattice vectors, we can rewrite the ferromagnetic triangular lattice problem of \eq{eq: 5} by combining \eq{Seq:4} and \eq{Seq:5} as
\begin{equation}
    \begin{split}
    \hat{\mathcal{H}}_{\text{FM}}=&\begin{pmatrix} \hat{x}_{0,+} \\ \hat{x}^+_{\text{G},l} \\ \hat{x}^+_{\text{G},t} \end{pmatrix}^{\dagger} \begin{pmatrix} \epsilon  & \frac{\sqrt{3}(\gamma+\epsilon)}{3} & \frac{\sqrt{3}(\gamma+\epsilon)}{3} \\ \frac{\sqrt{3}(\gamma+\epsilon)}{3} & \lambda_{\textbf{G},l}+\frac{\gamma+3\epsilon}{6}& \frac{3\gamma+\epsilon}{6}+\epsilon \\ \frac{\sqrt{3}(\gamma+\epsilon)}{3} & \frac{3\gamma+\epsilon}{6}+\epsilon & \lambda_{\textbf{G},t}+\frac{\gamma+3\epsilon}{6} \end{pmatrix}\begin{pmatrix} \hat{x}_{0,+} \\ \hat{x}^+_{\text{G},l} \\ \hat{x}^+_{\text{G},t} \end{pmatrix} + \\
    +&\begin{pmatrix} \hat{x}_{0,-} \\ \hat{x}^-_{\text{G},l} \\ \hat{x}^-_{\text{G},t} \end{pmatrix}^{\dagger} \begin{pmatrix} -\epsilon  & \frac{\sqrt{3}(\gamma-\epsilon)}{3} & \frac{\sqrt{3}(\epsilon-\gamma)}{3} \\ \frac{\sqrt{3}(\gamma-\epsilon)}{3} & \lambda_{\textbf{G},l}+\frac{\gamma-3\epsilon}{6} & \frac{\epsilon-3\gamma}{6}+\epsilon \\ \frac{\sqrt{3}(\epsilon-\gamma)}{3} & \frac{\epsilon-3\gamma}{6}+\epsilon & \lambda_{\textbf{G},t}+\frac{\gamma-3\epsilon}{6} \end{pmatrix} \begin{pmatrix} \hat{x}_{0,-} \\ \hat{x}^-_{\text{G},l} \\ \hat{x}^-_{\text{G},t} \end{pmatrix},
    \label{Seq:6}
    \end{split}
\end{equation}
where $\gamma=n_e(\lambda+\mu)/2$ and $\epsilon=n_e(\lambda-\mu)/2$ and $n_e$ is the average electron density.


\section{Ferromagnetic and antiferromagnetic honeycomb lattices}

The model developed for a $C_6$ magnetic potential can be extended to $C_3$ potentials, typical of honeycomb structures. Since the symmetry of the system has been reduced and is now invariant under rotations of multiples of $2\pi/3$, we expect the previously $\sigma^+$ and $\sigma^-$ subspaces to contain additional states. A simple analysis similar to that of \eq{eq:3} shows that in general there are four finite momentum states coupling to each zero momentum state, resulting in five-dimensional subspaces. Intuitively, each state of the symmetry basis defined in \eq{eq:3} splits into two states, containing phased superpositions that fulfil the $C_3$ symmetry. In general, the Fourier components of the first harmonics of a $C_3$ symmetric potential can be parametrised as $B_{\textbf{G}}=|B_\textbf{G}|e^{i\varphi_{\textbf{G}}}/2$. The components corresponding to lattice vectors \textbf{G} that are connected via a $C_3$ rotation must be equal. Moreover, the magnetic field is real, which imposes that $B_{\textbf{G}}=B_{-\textbf{G}}$. These conditions enforce that a $\pi/3$ rotation is equivalent to a complex conjugation of the original Fourier component. That is, there are two sets of three Fourier components connected via $2\pi/3$ rotations, given by $|B_\textbf{G}|e^{i\varphi}/2$ and $|B_\textbf{G}|e^{-i\varphi}/2$ respectively. Following the procedure detailed in the $C_6$ case, we can define the symmetry basis as
\begin{equation}
\begin{split}
    \hat{x}^{\pm}_{1,\nu}=\frac{1}{\sqrt{3}}(\hat{x}_{\textbf{G}1+\textbf{G}2,\nu}+e^{\pm2\pi/3}\hat{x}_{-\textbf{G}2,\nu}+e^{\pm4\pi/3}\hat{x}_{-\textbf{G}1,\nu}),\\
    \hat{x}^{\pm}_{2,\nu}=\frac{1}{\sqrt{3}}(e^{\pm i \pi/3}\hat{x}_{\textbf{G}1,\nu}+e^{\pm\pi}\hat{x}_{-\textbf{G}1-\textbf{G}2,\nu}+e^{\pm5\pi/3}\hat{x}_{\textbf{G}2,\nu}),
\end{split}
\end{equation}
with $\nu={l,t}$ and compute the matrix elements of the Hamiltonian for a purely periodic identity (top sign) and magnetic potential (bottom sign), which results in
\small
\begin{equation}
\begin{split}
    \hat{\mathcal{H}}=&\begin{pmatrix} \hat{x}_{0,+} \\ \hat{x}^+_{1,l} \\ \hat{x}^+_{2,l} \\ \hat{x}^+_{1,t} \\ \hat{x}^+_{2,t} \end{pmatrix}^{\dagger}
    \begin{pmatrix} 0 & \sqrt{\frac{3}{8}}\Delta e^{i\varphi} & \sqrt{\frac{3}{8}}\Delta e^{-i\varphi} & \sqrt{\frac{3}{8}}\Delta e^{i\varphi} & \sqrt{\frac{3}{8}}\Delta e^{-i\varphi} \\ \sqrt{\frac{3}{8}}\Delta e^{-i\varphi} & \lambda_{\boldsymbol{G},l} & \frac{2\mp1}{4}\Delta e^{i\varphi} & 0 & \frac{2\pm1}{4}\Delta e^{i\varphi} \\    \sqrt{\frac{3}{8}}\Delta e^{i\varphi} & \frac{2\mp1}{4}\Delta e^{-i\varphi} & \lambda_{\boldsymbol{G},l} & \frac{2\pm1}{4}\Delta e^{-i\varphi} & 0\\    \sqrt{\frac{3}{8}}\Delta e^{-i\varphi} &  0 & \frac{2\pm1}{4}\Delta e^{i\varphi} &  \lambda_{\boldsymbol{G},t} & \frac{2\mp1}{4}\Delta e^{i\varphi}\\ \sqrt{\frac{3}{8}}\Delta e^{i\varphi} & \frac{2\pm1}{4}\Delta e^{-i\varphi} & 0 & \frac{2\mp1}{4}\Delta e^{-i\varphi} & \lambda_{\boldsymbol{G},t} \end{pmatrix}\begin{pmatrix} \hat{x}_{0,+} \\ \hat{x}^+_{1,l} \\    \hat{x}^+_{2,l} \\ \hat{x}^+_{1,t} \\ \hat{x}^+_{2,t} \end{pmatrix} + \\ &    \begin{pmatrix} \hat{x}_{0,-} \\ \hat{x}^-_{1,l} \\ \hat{x}^-_{2,l} \\ \hat{x}^-_{1,t} \\ \hat{x}^-_{2,t} \end{pmatrix}^{\dagger}
    \begin{pmatrix} 0 & \pm\sqrt{\frac{3}{8}}\Delta e^{i\varphi} & \pm\sqrt{\frac{3}{8}}\Delta e^{-i\varphi} & \mp\sqrt{\frac{3}{8}}\Delta e^{i\varphi} & \mp\sqrt{\frac{3}{8}}\Delta e^{-i\varphi} \\ \pm\sqrt{\frac{3}{8}}\Delta e^{-i\varphi} & \lambda_{\boldsymbol{G},l} & \frac{-1\pm2}{4}\Delta e^{i\varphi} & 0 & \frac{-1\mp2}{4}\Delta e^{i\varphi} \\    \pm\sqrt{\frac{3}{8}}\Delta e^{i\varphi} & \frac{-1\pm2}{4}\Delta e^{-i\varphi} & \lambda_{\boldsymbol{G},l} & \frac{-1\mp2}{4}\Delta e^{-i\varphi} & 0\\\mp\sqrt{\frac{3}{8}}\Delta e^{-i\varphi} &  0 & \frac{-1\mp2}{4}\Delta e^{i\varphi} &  \lambda_{\boldsymbol{G},t} & \frac{-1\pm2}{4} \Delta e^{i\varphi}\\ \mp\sqrt{\frac{3}{8}}\Delta e^{i\varphi} & \frac{-1\mp2}{4}\Delta e^{-i\varphi} & 0 & \frac{-1\pm2}{4}\Delta e^{-i\varphi} & \lambda_{\boldsymbol{G},t} \end{pmatrix}\begin{pmatrix} \hat{x}_{0,-} \\ \hat{x}^-_{1,l} \\    \hat{x}^-_{2,l} \\ \hat{x}^-_{1,t} \\ \hat{x}^-_{2,t} \end{pmatrix}.
\end{split}
\end{equation}
\normalsize
We rewrite the Hamiltonian in a more convenient basis containing phased superposition of $\hat{x}^{\pm}_{1,\nu}$ and $\hat{x}^{\pm}_{2,\nu}$, given by the unitary transformation
\small
\begin{equation}
    \begin{pmatrix} \hat{x}_{0,-} \\ \hat{x}^-_{s,l} \\    \hat{x}^-_{a,l} \\ \hat{x}^-_{s,t} \\ \hat{x}^-_{a,t} \end{pmatrix}=\begin{pmatrix} 1 & 0 & 0 & 0 & 0\\ 0 & \frac{e^{i\varphi/2}}{\sqrt{2}} & \frac{e^{i\varphi/2}}{\sqrt{2}} & 0 & 0 \\ 0 & \frac{e^{-i\varphi/2}}{\sqrt{2}} & -\frac{e^{-i\varphi/2}}{\sqrt{2}} & 0 & 0 \\ 0&0&0& \frac{e^{i\varphi/2}}{\sqrt{2}} & \frac{e^{i\varphi/2}}{\sqrt{2}} \\ 0&0&0& \frac{e^{-i\varphi/2}}{\sqrt{2}} & -\frac{e^{-i\varphi/2}}{\sqrt{2}} \end{pmatrix} \begin{pmatrix} \hat{x}_{0,-} \\ \hat{x}^-_{1,l} \\    \hat{x}^-_{2,l} \\ \hat{x}^-_{1,t} \\ \hat{x}^-_{2,t} \end{pmatrix},
\end{equation}
\normalsize
where $s$ and $a$ are the new basis labels, corresponding to symmetric and antisymmetric superpositions respectively in the case of $\varphi=0$. Performing the basis transformation and rearranging its elements, we obtain
\small
\begin{equation}
\begin{split}
    \hat{\mathcal{H}}=&\begin{pmatrix} \hat{x}_{0,+} \\ \hat{x}^+_{s,l} \\ \hat{x}^+_{s,t} \\ \hat{x}^+_{a,l} \\ \hat{x}^+_{a,t} \end{pmatrix}^{\dagger}
    \begin{pmatrix} 0 & \frac{\sqrt{3}\Delta}{2} \cos{3\varphi/2} & i\frac{\sqrt{3}\Delta}{2} \cos{3\varphi/2} & i\frac{\sqrt{3}\Delta}{2}\sin{3\varphi/2} & i\frac{\sqrt{3}\Delta}{2} \sin{3\varphi/2} \\ \frac{\sqrt{3}\Delta}{2} \cos{3\varphi/2} &  \lambda_{\boldsymbol{G},l} +\frac{2\mp1}{4}\Delta & \frac{2\pm1}{4}\Delta & 0 & 0 \\ \frac{\sqrt{3}\Delta}{2} \cos{3\varphi/2} & \frac{2\pm1}{4}\Delta & \lambda_{\boldsymbol{G},l} +\frac{2\mp1}{4}\Delta & 0 & 0 \\ -i\frac{\sqrt{3}\Delta}{2} \sin{3\varphi/2} & 0 & 0 & \lambda_{\boldsymbol{G},l} - \frac{2\mp1}{4}\Delta & -\frac{2\pm1}{4}\Delta\\ -i\frac{\sqrt{3}\Delta}{2} \sin{3\varphi/2} & 0 & 0 & -\frac{2\pm1}{4}\Delta & \lambda_{\boldsymbol{G},l} - \frac{2\mp1}{4}\Delta  \end{pmatrix}\begin{pmatrix} \hat{x}_{0,+} \\ \hat{x}^+_{s,l} \\ \hat{x}^+_{s,t} \\ \hat{x}^+_{a,l} \\ \hat{x}^+_{a,t}\end{pmatrix} + \\ &    \begin{pmatrix} \hat{x}_{0,-} \\ \hat{x}^-_{s,l} \\ \hat{x}^-_{s,t} \\ \hat{x}^-_{a,l} \\ \hat{x}^-_{a,t} \end{pmatrix}^{\dagger}
    \begin{pmatrix} 0 & \pm\frac{\sqrt{3}\Delta}{2} \cos{3\varphi/2} & \mp\frac{\sqrt{3}\Delta}{2} \cos{3\varphi/2} & \pm i\frac{\sqrt{3}\Delta}{2}\sin{3\varphi/2} & \mp i\frac{\sqrt{3}\Delta}{2} \sin{3\varphi/2} \\ \pm\frac{\sqrt{3}\Delta}{2} \cos{3\varphi/2} &  \lambda_{\boldsymbol{G},l} -\frac{2\mp1}{4}\Delta & \frac{2\pm1}{4}\Delta & 0 & 0 \\ \mp\frac{\sqrt{3}\Delta}{2} \cos{3\varphi/2} & \frac{2\pm1}{4}\Delta & \lambda_{\boldsymbol{G},l} -\frac{2\mp1}{4}\Delta & 0 & 0 \\ \mp i\frac{\sqrt{3}\Delta}{2} \sin{3\varphi/2} & 0 & 0 & \lambda_{\boldsymbol{G},l} + \frac{2\mp1}{4}\Delta & -\frac{2\pm1}{4}\Delta\\ \pm i\frac{\sqrt{3}\Delta}{2} \sin{3\varphi/2} & 0 & 0 & - \frac{2\pm1}{4}\Delta & \lambda_{\boldsymbol{G},l} + \frac{2\mp1}{4}\Delta  \end{pmatrix}\begin{pmatrix} \hat{x}_{0,-} \\ \hat{x}^-_{s,l} \\ \hat{x}^-_{s,t} \\ \hat{x}^-_{a,l} \\ \hat{x}^-_{a,t}\end{pmatrix}.
\end{split}
\label{Seq:10}
\end{equation}
\normalsize
In general, both symmetric and antisymmetric subspaces couple to the zero momentum excitons and therefore four Umklapp resonances appear for each $\sigma^+$ and $\sigma^-$ subspace. However, for certain values of $\varphi$, the symmetric or antisymmetric superposition excitons decouple from the zero momentum excitons and become dark. In particular, it is clear that for $\varphi=0$ we recover the results obtained for the $C_6$ case, since the antisymmetric subspace no longer interacts with the zero momentum excitons and the new basis coincide with the states defined in \eq{eq:3}. 

\begin{figure}
    \centering
    \includegraphics[scale=0.4]{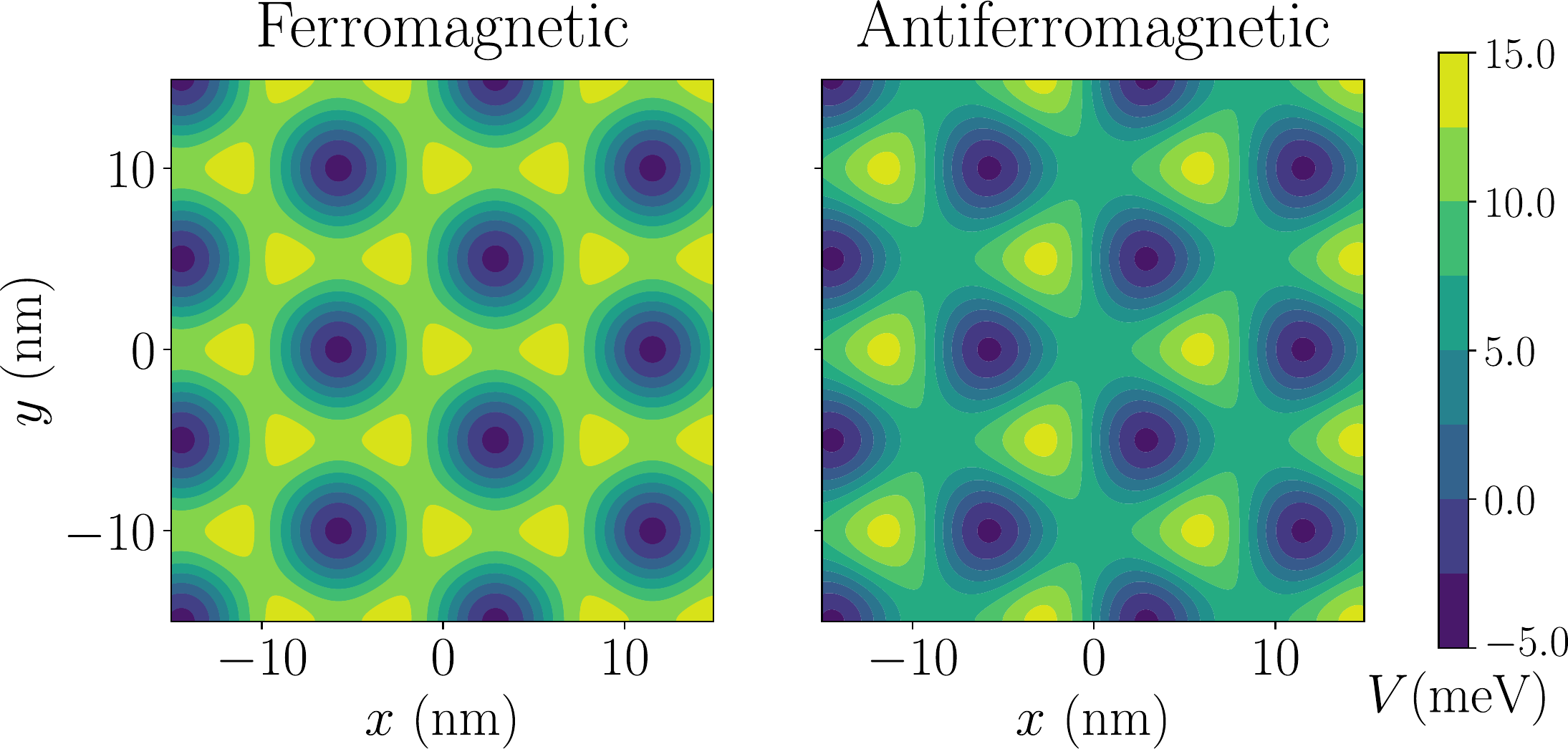}
    \caption{Effective potential $V$ felt by $\sigma^+$ excitons when interacting with a honeycomb lattice of electrons presenting ferromagnetic and antiferromagnetic ordering. In both plots we use $n_e \lambda=\SI{2}{\milli\electronvolt}$ and $n_e \mu=\SI{4}{\milli\electronvolt}$. In the ferromagnetic scenario, the particular value $\varphi=\pi/3$ for both identity and magnetic potential imposes a $C_6$ symmetry. In the antiferromagnetic scenario, the addition of the identity and magnetic potential characterized by $\varphi=\pi/3$ and $\varphi=-\pi/6$ results in a generic $C_3$ potential.}
    \label{Sfig:1}
\end{figure}

\medskip

Consider a honeycomb lattice with triangular sublattices labeled $A$ and $B$. We denote as $n_A(\textbf{r})$ and $n_B(\textbf{r})$ the density envelopes corresponding to the $A$ and $B$ sublattice respectively. Given $n_A(\textbf{r})=\frac{2n_e}{3}\big{\{}\cos{\big{[}\text{G}\big{(}\frac{x}{2}+\frac{\sqrt{3}y}{2}\big{)}\big{]}}+\cos{\big{[}\text{G}\big{(}\frac{x}{2}-\frac{\sqrt{3}y}{2}\big{)}\big{]}}+\cos{(\text{G}x)}+\frac{3}{2} \big{\}}$, the envelope for $B$ is given by $n_B(\textbf{r})=n_A(\textbf{r}-\frac{a}{\sqrt{3}}\textbf{e}_x)$. 

\medskip

In the ferromagnetic (FM) honeycomb lattice, spins of both sublattices point in the same direction and the spin density reads $n_e^{\text{FM}}(\textbf{r})=n_A(\textbf{r})+n_B(\textbf{r})$. Since only one spin species is present, the interaction Hamiltionian presents the same structure as \eq{eq: 5} of the main text and the angle $\varphi$ characterising the $C_3$ potential is given by $\tan^{-1}(\frac{\sin{2\pi/3}}{1+\cos{2\pi/3}})=\pi/3$. If we take a closer look at \eq{Seq:10}, we realize that for $\varphi=\pi/3$ two Umklapp states in each sector become dark and recover a triangular lattice signature, i.e., two split Umklapp resonances, one belonging to each subspace. This can be intuitively understood by considering \fig{Sfig:1} where we plot the potential excitons feel when interacting with a FM honeycomb lattice; such potential presents a $C_6$ symmetry and therefore results in two bright peaks and two dark ones.

\medskip

In the antiferromagnetic (AFM) honeycomb lattice on the other hand, spins of different sublattices point in opposite directions. The Hamiltonian for this scneario can be recast as
\begin{equation}
\begin{split}
    \hat{\mathcal{H}}_{\text{int}}^{\text{AFM}}=&\begin{pmatrix}  \hat{x}_{+}(\textbf{r})\\ \hat{x}_{-}(\textbf{r}) \end{pmatrix} ^\dagger \begin{pmatrix}\lambda\; n_A(\textbf{r})+\mu\;n_B(\textbf{r}) &0\\0&\mu\; n_A(\textbf{r})+\lambda\;n_B(\textbf{r}) \end{pmatrix} \begin{pmatrix}  \hat{x}_{+}(\textbf{r})\\ \hat{x}_{-}(\textbf{r})
    \end{pmatrix}\\=& \begin{pmatrix}  \hat{x}_{+}(\textbf{r})\\ \hat{x}_{-}(\textbf{r}) \end{pmatrix} ^\dagger \bigg{(} \frac{\lambda+\mu}{2}\big{[}n_A(\textbf{r})+n_B(\textbf{r})\big{]}\:\mathbb{1}+\frac{\lambda-\mu}{2}\big{[}n_A(\textbf{r})-n_B(\textbf{r})\big{]}\:\sigma_z \bigg{)} \begin{pmatrix}  \hat{x}_{+}(\textbf{r})\\ \hat{x}_{-}(\textbf{r}) \end{pmatrix},
\end{split}
\label{Seq:11}
\end{equation}
where we can directly read the identity and magnetic potential in terms of the triangular lattice distributions. The density profile of the magnetic term is therefore given by $n_e^{\text{AFM}}(\textbf{r})=n_A(\textbf{r})-n_B(\textbf{r})$ and the characteristic phase by  $\tan^{-1}(\frac{-\sin{2\pi/3}}{1-\cos{2\pi/3}})=-\pi/6$. Hence, when studying the AFM lattice, the identity and magnetic potential are characterised by $\varphi=\pi/3$ and $\varphi=-\pi/6$ respectively. 

In order to apply \eq{Seq:10} to the FM and AFM honeycomb lattices, one needs to identify $\Delta=\frac{\sqrt{2}}{3}n_e(\lambda+\mu)$ 
for the periodic identity term and $\Delta=\frac{\sqrt{2}}{3}n_e(\lambda-\mu)$ for the periodic magnetic term. Note that \eq{Seq:10} has been derived for identity and magnetic potentials with a vanishing average. Thus, we further introduce a constant energy shift given by $n_e(\lambda+\mu)$ to take into account the net electron density for both scenarios, and a constant magnetic field given by $n_e(\lambda-\mu)$ for the FM case, to account for the net magnetization.


\section{Twisted bilayer of CrI\texorpdfstring{$_3$}{3}} 

{\renewcommand{\arraystretch}{2.0}
\begin{figure}
\begin{floatrow}
\capbtabbox{%
 \begin{tabular}{cc} \hline
  \textbf{Harmonic} & \textbf{Coefficient} \\ \hline
  0th & -0.215 \\ \hline
  1st & 0.055+0.099$i$ \\ \hline
  2nd & -0.205-0.147$i$ \\ \hline
  3rd & -0.084+0.159$i$ \\ \hline
  \end{tabular}
}{%
  \caption{Coefficients of the first four Fourier components fitted to the spin distribution that arises from the interlayer exchange function in Ref.~\cite{Sivadas_2018}.}
 \label{Tab: Harmonics}
}

\ffigbox{%
    \includegraphics[scale=1.1]{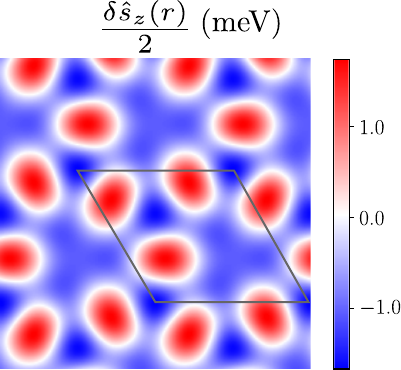}}
    {%
  \caption{Plot of the first four Fourier components of the spatial distribution of the periodic exchange interaction strength for $\delta=\SI{2}{meV}$. The grey parallelogram encloses the unit cell.}
  \label{Sfig:2}%
}

\end{floatrow}
\end{figure}
}

In this section, we briefly review the work on moiré magnets in twisted bilayers of CrI$_3$ presented in Ref.~\cite{Balents_2020}. Consider two monolayers of CrI$_3$, which are rotated with respect to each other by an angle $\theta$. The spins within each layer interact by means of an intra-layer exchange coupling $J$ and the two layers are coupled by an inter-layer exchange coupling $J'$. We assume that $J \gg J'$ allowing for a continuum representation of the magnetism of each layer. Hence, each layer can be entirely characterised by a magnetisation vector. Under these considerations, the ground state of the system can be found as the variational minimum of the classical Hamiltonian
\begin{equation}
    \mathcal{H}_{cl}=\sum_{j}\bigg{[} \frac{\rho}{2}(\nabla \boldsymbol{M}_j)^2-d(M^z_j)^2 \bigg{]} -J'\Phi(\boldsymbol{r})\boldsymbol{M}_1\cdot \boldsymbol{M}_2,
    \label{Seq:12}
\end{equation}
where $\boldsymbol{M}_j$ is the normalised magnetisation vector of the $j$th layer with $j\in\{1,2\}$, which can be considered, without any loss of generality, to lie in the $x$--$z$ plane: $\boldsymbol{M}_j=\sin{\phi_j}\textbf{e}_x+\cos{\phi_j}\textbf{e}_z$. Here $\rho \sim J$ is the spin stiffness, $d$ is the uniaxial anisotropy in $z$ direction, and $J'\Phi(\boldsymbol{r})$ is the interlayer exchange function. The relative displacement between the two layers is periodic for twisted bilayers, which implies that the interlayer exchange function can be expanded in the reciprocal lattice vectors of the moiré superlattice: $\boldsymbol{q}_a=\theta\: \textbf{e}_z\times \boldsymbol{b}_a$, with $\boldsymbol{b}_a$ the reciprocal lattice vectors of CrI$_3$. The variational minimization fixes $ \phi_1$ and $\phi_2$ as a function of the dimensionless parameters $\tilde{\alpha}\equiv2J'/\rho|\boldsymbol{q}_a|^2$ and $\tilde{\beta}\equiv2d/\rho|\boldsymbol{q}_a|^2$. For typical values of CrI$_3$ and small twisting angles $\theta\lesssim5^{\circ}$, both $\tilde{\alpha},\tilde{\beta}\gg1$. In this regime, the arising magnetic landscape of one layer presents negligible spatial modulation while the other follows the sign of $\Phi(\boldsymbol{r})$ except for small domains where $\Phi(\boldsymbol{r})=0$. This results in spins pointing up in the regions where the interlayer exchange is ferromagnetic and pointing down in the antiferromagnetic regions, resulting in a periodic modulation of $\hat{s}_z(\boldsymbol{r})$. The precise distribution for the ferromagnetic and antiferromagnetic interlayer regions is described by $J'\Phi(\boldsymbol{r})$, obtained via density field calculations in Ref.~\cite{Sivadas_2018}. Using the interlayer exchange function, we have estimated the arising spin distribution and have extracted its first four Fourier components, presented in \tab{Tab: Harmonics}. The periodic exchange interaction, which acts as effective magnetic field on the excitons is presented in \fig{Sfig:2}. For completeness we present the zero momentum excitonic spectral function in \figc{Sfig:3}{a}. As discussed in the main text, one of the Umklapp peaks falls within the linewidth of the main resonance and is cannot be optically resolved. However, the second Umklapp peak splits notably and presents an experimentally accessible oscillator strength on the order of 2\,\%.


\section{Estimation of interaction parameters \texorpdfstring{$\lambda$}{l} and \texorpdfstring{$\mu$}{m}}

\begin{figure}
\centering
\includegraphics{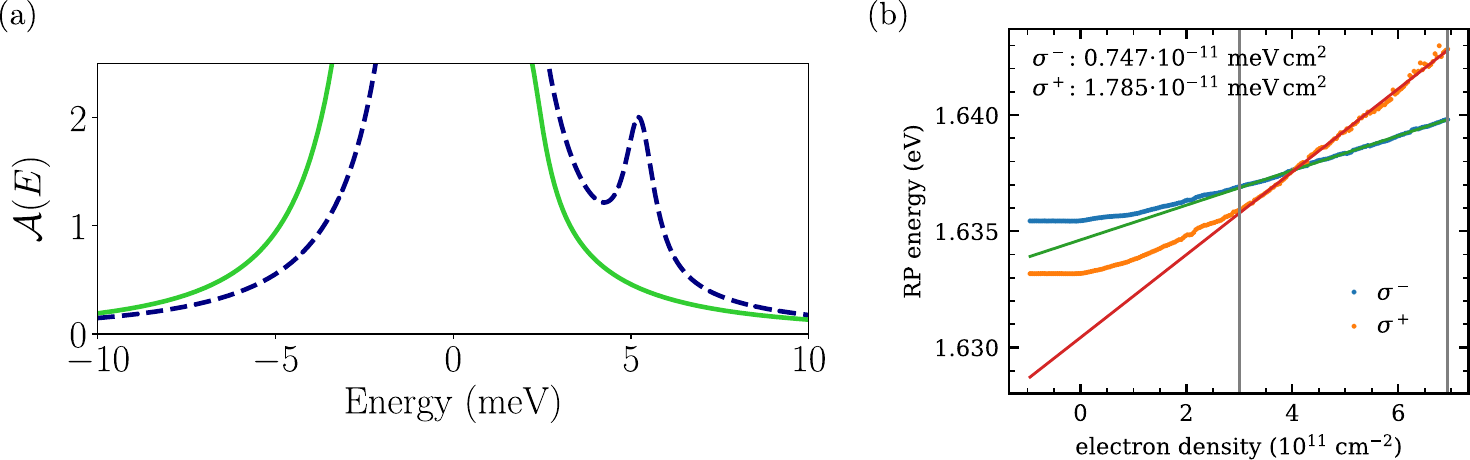}
\caption{\fc{a} Excitonic spectral function resulting from a proximity coupling to the CrI$_3$ bilayer of $\delta=\SI{3}{\milli\electronvolt}$ and a moiré periodicity of \SI{20}{\nano\meter}. In solid green (dashed navy blue) the spectral function of the $\sigma^-$ ($\sigma^+$) subspace. The arising Umklapp peak provides evidence of the presence of a periodic magnetisation in the twisted bilayer.  \fc{b} Fitted repulsive polaron energies in K and K' valley as function of electron density at $B=\SI{8.5}{\tesla}$. The linear fits
to the slope at intermediate densities correspond to the parameters $\lambda$ and $\mu$ in the main text.}
\label{Sfig:3}
\end{figure}

Realistic values for the inter- and intravalley interaction parameters $\lambda$ and $\mu$ were obtained from
polarization-resolved reflection spectra of a MoSe$_2$ monolayer. Linearly polarized light from a superluminescent LED
with a center wavelength of \SI{760}{\nano\meter} and a width of \SI{40}{\nano\meter} was focused on a gate-tunable MoSe$_2$ monolayer encapsulated
between two approximately \SI{75}{\nano\meter} thick hexagonal boron nitride flakes. The right- and left-hand circularly polarized components
of the reflected light were separated and detected simultaneously with a spectrometer. The spectrum was measured as a function of gate-voltage
in a constant magnetic field of $B = \SI{8.5}{\tesla}$, which ensured almost full valley polarization at the relevant electron densities~\cite{Back17}.
The spectra were normalized by the spectrum taken on a spot in the heterostructure without the MoSe$_2$. The repulsive polaron resonances were fitted
with a complex Lorentzian $f(\lambda) = \mathcal{R}(A \exp(i\varphi) / (\Gamma - i (\lambda - \lambda_0))) + c$ in order to extract the energy of the repulsive
polaron as function of density. The paramters $\mu$ and $\lambda$ were extracted from the linear slope of the energy vs.\ density in the two
polarziations in the density range between \SI{3e11}{\per\centi\meter\squared} and \SI{7e11}{\per\centi\meter\squared}. Fitted repulsive polaron
energies as well as the extracted slopes are shown in \figc{Sfig:3}{b}.


\begin{thebibliography}{37}%
\makeatletter
\providecommand \@ifxundefined [1]{%
 \@ifx{#1\undefined}
}%
\providecommand \@ifnum [1]{%
 \ifnum #1\expandafter \@firstoftwo
 \else \expandafter \@secondoftwo
 \fi
}%
\providecommand \@ifx [1]{%
 \ifx #1\expandafter \@firstoftwo
 \else \expandafter \@secondoftwo
 \fi
}%
\providecommand \natexlab [1]{#1}%
\providecommand \enquote  [1]{``#1''}%
\providecommand \bibnamefont  [1]{#1}%
\providecommand \bibfnamefont [1]{#1}%
\providecommand \citenamefont [1]{#1}%
\providecommand \href@noop [0]{\@secondoftwo}%
\providecommand \href [0]{\begingroup \@sanitize@url \@href}%
\providecommand \@href[1]{\@@startlink{#1}\@@href}%
\providecommand \@@href[1]{\endgroup#1\@@endlink}%
\providecommand \@sanitize@url [0]{\catcode `\\12\catcode `\$12\catcode
  `\&12\catcode `\#12\catcode `\^12\catcode `\_12\catcode `\%12\relax}%
\providecommand \@@startlink[1]{}%
\providecommand \@@endlink[0]{}%
\providecommand \url  [0]{\begingroup\@sanitize@url \@url }%
\providecommand \@url [1]{\endgroup\@href {#1}{\urlprefix }}%
\providecommand \urlprefix  [0]{URL }%
\providecommand \Eprint [0]{\href }%
\providecommand \doibase [0]{https://doi.org/}%
\providecommand \selectlanguage [0]{\@gobble}%
\providecommand \bibinfo  [0]{\@secondoftwo}%
\providecommand \bibfield  [0]{\@secondoftwo}%
\providecommand \translation [1]{[#1]}%
\providecommand \BibitemOpen [0]{}%
\providecommand \bibitemStop [0]{}%
\providecommand \bibitemNoStop [0]{.\EOS\space}%
\providecommand \EOS [0]{\spacefactor3000\relax}%
\providecommand \BibitemShut  [1]{\csname bibitem#1\endcsname}%
\let\auto@bib@innerbib\@empty
\bibitem [{\citenamefont {Cao}\ \emph {et~al.}(2018)\citenamefont {Cao},
  \citenamefont {Fatemi}, \citenamefont {Fang}, \citenamefont {Watanabe},
  \citenamefont {Taniguchi}, \citenamefont {Kaxiras},\ and\ \citenamefont
  {Jarillo-Herrero}}]{Jarrillo-Herrero2018}%
  \BibitemOpen
  \bibfield  {author} {\bibinfo {author} {\bibfnamefont {Y.}~\bibnamefont
  {Cao}}, \bibinfo {author} {\bibfnamefont {V.}~\bibnamefont {Fatemi}},
  \bibinfo {author} {\bibfnamefont {S.}~\bibnamefont {Fang}}, \bibinfo {author}
  {\bibfnamefont {K.}~\bibnamefont {Watanabe}}, \bibinfo {author}
  {\bibfnamefont {T.}~\bibnamefont {Taniguchi}}, \bibinfo {author}
  {\bibfnamefont {E.}~\bibnamefont {Kaxiras}},\ and\ \bibinfo {author}
  {\bibfnamefont {P.}~\bibnamefont {Jarillo-Herrero}},\ }\href
  {https://doi.org/10.1038/nature26160} {\bibfield  {journal} {\bibinfo
  {journal} {Nature}\ }\textbf {\bibinfo {volume} {556}},\ \bibinfo {pages}
  {43} (\bibinfo {year} {2018})}\BibitemShut {NoStop}%
\bibitem [{\citenamefont {Balents}\ \emph {et~al.}(2020)\citenamefont
  {Balents}, \citenamefont {Dean}, \citenamefont {Efetov},\ and\ \citenamefont
  {Young}}]{Young20}%
  \BibitemOpen
  \bibfield  {author} {\bibinfo {author} {\bibfnamefont {L.}~\bibnamefont
  {Balents}}, \bibinfo {author} {\bibfnamefont {C.~R.}\ \bibnamefont {Dean}},
  \bibinfo {author} {\bibfnamefont {D.~K.}\ \bibnamefont {Efetov}},\ and\
  \bibinfo {author} {\bibfnamefont {A.~F.}\ \bibnamefont {Young}},\ }\href
  {https://doi.org/10.1038/s41567-020-0906-9} {\bibfield  {journal} {\bibinfo
  {journal} {Nature Physics}\ }\textbf {\bibinfo {volume} {16}},\ \bibinfo
  {pages} {725} (\bibinfo {year} {2020})}\BibitemShut {NoStop}%
\bibitem [{\citenamefont {Lu}\ \emph {et~al.}(2019)\citenamefont {Lu},
  \citenamefont {Stepanov}, \citenamefont {Yang}, \citenamefont {Xie},
  \citenamefont {Aamir}, \citenamefont {Das}, \citenamefont {Urgell},
  \citenamefont {Watanabe}, \citenamefont {Taniguchi}, \citenamefont {Zhang},\
  and\ \citenamefont {et~al.}}]{Lu2019}%
  \BibitemOpen
  \bibfield  {author} {\bibinfo {author} {\bibfnamefont {X.}~\bibnamefont
  {Lu}}, \bibinfo {author} {\bibfnamefont {P.}~\bibnamefont {Stepanov}},
  \bibinfo {author} {\bibfnamefont {W.}~\bibnamefont {Yang}}, \bibinfo {author}
  {\bibfnamefont {M.}~\bibnamefont {Xie}}, \bibinfo {author} {\bibfnamefont
  {M.~A.}\ \bibnamefont {Aamir}}, \bibinfo {author} {\bibfnamefont
  {I.}~\bibnamefont {Das}}, \bibinfo {author} {\bibfnamefont {C.}~\bibnamefont
  {Urgell}}, \bibinfo {author} {\bibfnamefont {K.}~\bibnamefont {Watanabe}},
  \bibinfo {author} {\bibfnamefont {T.}~\bibnamefont {Taniguchi}}, \bibinfo
  {author} {\bibfnamefont {G.}~\bibnamefont {Zhang}},\ and\ \bibinfo {author}
  {\bibnamefont {et~al.}},\ }\href {https://doi.org/10.1038/s41586-019-1695-0}
  {\bibfield  {journal} {\bibinfo  {journal} {Nature}\ }\textbf {\bibinfo
  {volume} {574}},\ \bibinfo {pages} {653–657} (\bibinfo {year}
  {2019})}\BibitemShut {NoStop}%
\bibitem [{\citenamefont {Regan}\ \emph {et~al.}(2020)\citenamefont {Regan},
  \citenamefont {Wang}, \citenamefont {Jin}, \citenamefont {Bakti~Utama},
  \citenamefont {Gao}, \citenamefont {Wei}, \citenamefont {Zhao}, \citenamefont
  {Zhao}, \citenamefont {Zhang}, \citenamefont {Yumigeta}, \citenamefont
  {Blei}, \citenamefont {Carlstr{\"o}m}, \citenamefont {Watanabe},
  \citenamefont {Taniguchi}, \citenamefont {Tongay}, \citenamefont {Crommie},
  \citenamefont {Zettl},\ and\ \citenamefont {Wang}}]{Regan2020}%
  \BibitemOpen
  \bibfield  {author} {\bibinfo {author} {\bibfnamefont {E.~C.}\ \bibnamefont
  {Regan}}, \bibinfo {author} {\bibfnamefont {D.}~\bibnamefont {Wang}},
  \bibinfo {author} {\bibfnamefont {C.}~\bibnamefont {Jin}}, \bibinfo {author}
  {\bibfnamefont {M.~I.}\ \bibnamefont {Bakti~Utama}}, \bibinfo {author}
  {\bibfnamefont {B.}~\bibnamefont {Gao}}, \bibinfo {author} {\bibfnamefont
  {X.}~\bibnamefont {Wei}}, \bibinfo {author} {\bibfnamefont {S.}~\bibnamefont
  {Zhao}}, \bibinfo {author} {\bibfnamefont {W.}~\bibnamefont {Zhao}}, \bibinfo
  {author} {\bibfnamefont {Z.}~\bibnamefont {Zhang}}, \bibinfo {author}
  {\bibfnamefont {K.}~\bibnamefont {Yumigeta}}, \bibinfo {author}
  {\bibfnamefont {M.}~\bibnamefont {Blei}}, \bibinfo {author} {\bibfnamefont
  {J.~D.}\ \bibnamefont {Carlstr{\"o}m}}, \bibinfo {author} {\bibfnamefont
  {K.}~\bibnamefont {Watanabe}}, \bibinfo {author} {\bibfnamefont
  {T.}~\bibnamefont {Taniguchi}}, \bibinfo {author} {\bibfnamefont
  {S.}~\bibnamefont {Tongay}}, \bibinfo {author} {\bibfnamefont
  {M.}~\bibnamefont {Crommie}}, \bibinfo {author} {\bibfnamefont
  {A.}~\bibnamefont {Zettl}},\ and\ \bibinfo {author} {\bibfnamefont
  {F.}~\bibnamefont {Wang}},\ }\href
  {https://doi.org/10.1038/s41586-020-2092-4} {\bibfield  {journal} {\bibinfo
  {journal} {Nature}\ }\textbf {\bibinfo {volume} {579}},\ \bibinfo {pages}
  {359} (\bibinfo {year} {2020})}\BibitemShut {NoStop}%
\bibitem [{\citenamefont {Tang}\ \emph {et~al.}(2020)\citenamefont {Tang},
  \citenamefont {Li}, \citenamefont {Li}, \citenamefont {Xu}, \citenamefont
  {Liu}, \citenamefont {Barmak}, \citenamefont {Watanabe}, \citenamefont
  {Taniguchi}, \citenamefont {MacDonald}, \citenamefont {Shan},\ and\
  \citenamefont {Mak}}]{Tang2020}%
  \BibitemOpen
  \bibfield  {author} {\bibinfo {author} {\bibfnamefont {Y.}~\bibnamefont
  {Tang}}, \bibinfo {author} {\bibfnamefont {L.}~\bibnamefont {Li}}, \bibinfo
  {author} {\bibfnamefont {T.}~\bibnamefont {Li}}, \bibinfo {author}
  {\bibfnamefont {Y.}~\bibnamefont {Xu}}, \bibinfo {author} {\bibfnamefont
  {S.}~\bibnamefont {Liu}}, \bibinfo {author} {\bibfnamefont {K.}~\bibnamefont
  {Barmak}}, \bibinfo {author} {\bibfnamefont {K.}~\bibnamefont {Watanabe}},
  \bibinfo {author} {\bibfnamefont {T.}~\bibnamefont {Taniguchi}}, \bibinfo
  {author} {\bibfnamefont {A.~H.}\ \bibnamefont {MacDonald}}, \bibinfo {author}
  {\bibfnamefont {J.}~\bibnamefont {Shan}},\ and\ \bibinfo {author}
  {\bibfnamefont {K.~F.}\ \bibnamefont {Mak}},\ }\href
  {https://doi.org/10.1038/s41586-020-2085-3} {\bibfield  {journal} {\bibinfo
  {journal} {Nature}\ }\textbf {\bibinfo {volume} {579}},\ \bibinfo {pages}
  {353} (\bibinfo {year} {2020})}\BibitemShut {NoStop}%
\bibitem [{\citenamefont {Shimazaki}\ \emph {et~al.}(2020)\citenamefont
  {Shimazaki}, \citenamefont {Schwartz}, \citenamefont {Watanabe},
  \citenamefont {Taniguchi}, \citenamefont {Kroner},\ and\ \citenamefont
  {Imamo{\u g}lu}}]{Shimazaki2020}%
  \BibitemOpen
  \bibfield  {author} {\bibinfo {author} {\bibfnamefont {Y.}~\bibnamefont
  {Shimazaki}}, \bibinfo {author} {\bibfnamefont {I.}~\bibnamefont {Schwartz}},
  \bibinfo {author} {\bibfnamefont {K.}~\bibnamefont {Watanabe}}, \bibinfo
  {author} {\bibfnamefont {T.}~\bibnamefont {Taniguchi}}, \bibinfo {author}
  {\bibfnamefont {M.}~\bibnamefont {Kroner}},\ and\ \bibinfo {author}
  {\bibfnamefont {A.}~\bibnamefont {Imamo{\u g}lu}},\ }\href
  {https://doi.org/10.1038/s41586-020-2191-2} {\bibfield  {journal} {\bibinfo
  {journal} {Nature}\ }\textbf {\bibinfo {volume} {580}},\ \bibinfo {pages}
  {472} (\bibinfo {year} {2020})}\BibitemShut {NoStop}%
\bibitem [{\citenamefont {Xu}\ \emph {et~al.}(2020)\citenamefont {Xu},
  \citenamefont {Liu}, \citenamefont {Rhodes}, \citenamefont {Watanabe},
  \citenamefont {Taniguchi}, \citenamefont {Hone}, \citenamefont {Elser},
  \citenamefont {Mak},\ and\ \citenamefont {Shan}}]{Xu2020}%
  \BibitemOpen
  \bibfield  {author} {\bibinfo {author} {\bibfnamefont {Y.}~\bibnamefont
  {Xu}}, \bibinfo {author} {\bibfnamefont {S.}~\bibnamefont {Liu}}, \bibinfo
  {author} {\bibfnamefont {D.~A.}\ \bibnamefont {Rhodes}}, \bibinfo {author}
  {\bibfnamefont {K.}~\bibnamefont {Watanabe}}, \bibinfo {author}
  {\bibfnamefont {T.}~\bibnamefont {Taniguchi}}, \bibinfo {author}
  {\bibfnamefont {J.}~\bibnamefont {Hone}}, \bibinfo {author} {\bibfnamefont
  {V.}~\bibnamefont {Elser}}, \bibinfo {author} {\bibfnamefont {K.~F.}\
  \bibnamefont {Mak}},\ and\ \bibinfo {author} {\bibfnamefont {J.}~\bibnamefont
  {Shan}},\ }\href {https://doi.org/10.1038/s41586-020-2868-6} {\bibfield
  {journal} {\bibinfo  {journal} {Nature}\ }\textbf {\bibinfo {volume} {587}},\
  \bibinfo {pages} {214} (\bibinfo {year} {2020})}\BibitemShut {NoStop}%
\bibitem [{\citenamefont {Gao}\ \emph {et~al.}(2020)\citenamefont {Gao},
  \citenamefont {Li}, \citenamefont {Xin}, \citenamefont {Chen}, \citenamefont
  {Liu},\ and\ \citenamefont {Tian}}]{Gao2020}%
  \BibitemOpen
  \bibfield  {author} {\bibinfo {author} {\bibfnamefont {X.-G.}\ \bibnamefont
  {Gao}}, \bibinfo {author} {\bibfnamefont {X.-K.}\ \bibnamefont {Li}},
  \bibinfo {author} {\bibfnamefont {W.}~\bibnamefont {Xin}}, \bibinfo {author}
  {\bibfnamefont {X.-D.}\ \bibnamefont {Chen}}, \bibinfo {author}
  {\bibfnamefont {Z.-B.}\ \bibnamefont {Liu}},\ and\ \bibinfo {author}
  {\bibfnamefont {J.-G.}\ \bibnamefont {Tian}},\ }\href
  {https://doi.org/doi:10.1515/nanoph-2020-0024} {\bibfield  {journal}
  {\bibinfo  {journal} {Nanophotonics}\ }\textbf {\bibinfo {volume} {9}},\
  \bibinfo {pages} {1717} (\bibinfo {year} {2020})}\BibitemShut {NoStop}%
\bibitem [{\citenamefont {Li}\ \emph {et~al.}(2021)\citenamefont {Li},
  \citenamefont {Li}, \citenamefont {Regan}, \citenamefont {Wang},
  \citenamefont {Zhao}, \citenamefont {Kahn}, \citenamefont {Yumigeta},
  \citenamefont {Blei}, \citenamefont {Taniguchi}, \citenamefont {Watanabe},
  \citenamefont {Tongay}, \citenamefont {Zettl}, \citenamefont {Crommie},\ and\
  \citenamefont {Wang}}]{Li2021}%
  \BibitemOpen
  \bibfield  {author} {\bibinfo {author} {\bibfnamefont {H.}~\bibnamefont
  {Li}}, \bibinfo {author} {\bibfnamefont {S.}~\bibnamefont {Li}}, \bibinfo
  {author} {\bibfnamefont {E.~C.}\ \bibnamefont {Regan}}, \bibinfo {author}
  {\bibfnamefont {D.}~\bibnamefont {Wang}}, \bibinfo {author} {\bibfnamefont
  {W.}~\bibnamefont {Zhao}}, \bibinfo {author} {\bibfnamefont {S.}~\bibnamefont
  {Kahn}}, \bibinfo {author} {\bibfnamefont {K.}~\bibnamefont {Yumigeta}},
  \bibinfo {author} {\bibfnamefont {M.}~\bibnamefont {Blei}}, \bibinfo {author}
  {\bibfnamefont {T.}~\bibnamefont {Taniguchi}}, \bibinfo {author}
  {\bibfnamefont {K.}~\bibnamefont {Watanabe}}, \bibinfo {author}
  {\bibfnamefont {S.}~\bibnamefont {Tongay}}, \bibinfo {author} {\bibfnamefont
  {A.}~\bibnamefont {Zettl}}, \bibinfo {author} {\bibfnamefont {M.~F.}\
  \bibnamefont {Crommie}},\ and\ \bibinfo {author} {\bibfnamefont
  {F.}~\bibnamefont {Wang}},\ }\href
  {https://doi.org/10.1038/s41586-021-03874-9} {\bibfield  {journal} {\bibinfo
  {journal} {Nature}\ }\textbf {\bibinfo {volume} {597}},\ \bibinfo {pages}
  {650} (\bibinfo {year} {2021})}\BibitemShut {NoStop}%
\bibitem [{\citenamefont {Shimazaki}\ \emph {et~al.}(2021)\citenamefont
  {Shimazaki}, \citenamefont {Kuhlenkamp}, \citenamefont {Schwartz},
  \citenamefont {Smole\ifmmode~\acute{n}\else \'{n}\fi{}ski}, \citenamefont
  {Watanabe}, \citenamefont {Taniguchi}, \citenamefont {Kroner}, \citenamefont
  {Schmidt}, \citenamefont {Knap},\ and\ \citenamefont
  {Imamo\ifmmode~\breve{g}\else \u{g}\fi{}lu}}]{Shimazaki2021}%
  \BibitemOpen
  \bibfield  {author} {\bibinfo {author} {\bibfnamefont {Y.}~\bibnamefont
  {Shimazaki}}, \bibinfo {author} {\bibfnamefont {C.}~\bibnamefont
  {Kuhlenkamp}}, \bibinfo {author} {\bibfnamefont {I.}~\bibnamefont
  {Schwartz}}, \bibinfo {author} {\bibfnamefont {T.}~\bibnamefont
  {Smole\ifmmode~\acute{n}\else \'{n}\fi{}ski}}, \bibinfo {author}
  {\bibfnamefont {K.}~\bibnamefont {Watanabe}}, \bibinfo {author}
  {\bibfnamefont {T.}~\bibnamefont {Taniguchi}}, \bibinfo {author}
  {\bibfnamefont {M.}~\bibnamefont {Kroner}}, \bibinfo {author} {\bibfnamefont
  {R.}~\bibnamefont {Schmidt}}, \bibinfo {author} {\bibfnamefont
  {M.}~\bibnamefont {Knap}},\ and\ \bibinfo {author} {\bibfnamefont
  {A.}~\bibnamefont {Imamo\ifmmode~\breve{g}\else \u{g}\fi{}lu}},\ }\href
  {https://doi.org/10.1103/PhysRevX.11.021027} {\bibfield  {journal} {\bibinfo
  {journal} {Phys. Rev. X}\ }\textbf {\bibinfo {volume} {11}},\ \bibinfo
  {pages} {021027} (\bibinfo {year} {2021})}\BibitemShut {NoStop}%
\bibitem [{\citenamefont {Wang}\ \emph {et~al.}(2020)\citenamefont {Wang},
  \citenamefont {Shih}, \citenamefont {Ghiotto}, \citenamefont {Xian},
  \citenamefont {Rhodes}, \citenamefont {Tan}, \citenamefont {Claassen},
  \citenamefont {Kennes}, \citenamefont {Bai}, \citenamefont {Kim},
  \citenamefont {Watanabe}, \citenamefont {Taniguchi}, \citenamefont {Zhu},
  \citenamefont {Hone}, \citenamefont {Rubio}, \citenamefont {Pasupathy},\ and\
  \citenamefont {Dean}}]{Wang2020}%
  \BibitemOpen
  \bibfield  {author} {\bibinfo {author} {\bibfnamefont {L.}~\bibnamefont
  {Wang}}, \bibinfo {author} {\bibfnamefont {E.-M.}\ \bibnamefont {Shih}},
  \bibinfo {author} {\bibfnamefont {A.}~\bibnamefont {Ghiotto}}, \bibinfo
  {author} {\bibfnamefont {L.}~\bibnamefont {Xian}}, \bibinfo {author}
  {\bibfnamefont {D.~A.}\ \bibnamefont {Rhodes}}, \bibinfo {author}
  {\bibfnamefont {C.}~\bibnamefont {Tan}}, \bibinfo {author} {\bibfnamefont
  {M.}~\bibnamefont {Claassen}}, \bibinfo {author} {\bibfnamefont {D.~M.}\
  \bibnamefont {Kennes}}, \bibinfo {author} {\bibfnamefont {Y.}~\bibnamefont
  {Bai}}, \bibinfo {author} {\bibfnamefont {B.}~\bibnamefont {Kim}}, \bibinfo
  {author} {\bibfnamefont {K.}~\bibnamefont {Watanabe}}, \bibinfo {author}
  {\bibfnamefont {T.}~\bibnamefont {Taniguchi}}, \bibinfo {author}
  {\bibfnamefont {X.}~\bibnamefont {Zhu}}, \bibinfo {author} {\bibfnamefont
  {J.}~\bibnamefont {Hone}}, \bibinfo {author} {\bibfnamefont {A.}~\bibnamefont
  {Rubio}}, \bibinfo {author} {\bibfnamefont {A.~N.}\ \bibnamefont
  {Pasupathy}},\ and\ \bibinfo {author} {\bibfnamefont {C.~R.}\ \bibnamefont
  {Dean}},\ }\href {https://doi.org/10.1038/s41563-020-0708-6} {\bibfield
  {journal} {\bibinfo  {journal} {Nature Materials}\ }\textbf {\bibinfo
  {volume} {19}},\ \bibinfo {pages} {861} (\bibinfo {year} {2020})}\BibitemShut
  {NoStop}%
\bibitem [{\citenamefont {Zhang}\ \emph {et~al.}(2020)\citenamefont {Zhang},
  \citenamefont {Wang}, \citenamefont {Watanabe}, \citenamefont {Taniguchi},
  \citenamefont {Ueno}, \citenamefont {Tutuc},\ and\ \citenamefont
  {LeRoy}}]{Zhang2020}%
  \BibitemOpen
  \bibfield  {author} {\bibinfo {author} {\bibfnamefont {Z.}~\bibnamefont
  {Zhang}}, \bibinfo {author} {\bibfnamefont {Y.}~\bibnamefont {Wang}},
  \bibinfo {author} {\bibfnamefont {K.}~\bibnamefont {Watanabe}}, \bibinfo
  {author} {\bibfnamefont {T.}~\bibnamefont {Taniguchi}}, \bibinfo {author}
  {\bibfnamefont {K.}~\bibnamefont {Ueno}}, \bibinfo {author} {\bibfnamefont
  {E.}~\bibnamefont {Tutuc}},\ and\ \bibinfo {author} {\bibfnamefont {B.~J.}\
  \bibnamefont {LeRoy}},\ }\href {https://doi.org/10.1038/s41567-020-0958-x}
  {\bibfield  {journal} {\bibinfo  {journal} {Nature Physics}\ }\textbf
  {\bibinfo {volume} {16}},\ \bibinfo {pages} {1093} (\bibinfo {year}
  {2020})}\BibitemShut {NoStop}%
\bibitem [{\citenamefont {Hejazi}\ \emph {et~al.}(2020)\citenamefont {Hejazi},
  \citenamefont {Luo},\ and\ \citenamefont {Balents}}]{Balents_2020}%
  \BibitemOpen
  \bibfield  {author} {\bibinfo {author} {\bibfnamefont {K.}~\bibnamefont
  {Hejazi}}, \bibinfo {author} {\bibfnamefont {Z.-X.}\ \bibnamefont {Luo}},\
  and\ \bibinfo {author} {\bibfnamefont {L.}~\bibnamefont {Balents}},\
  }\href@noop {} {\bibinfo {title} {Moir\'e magnets}} (\bibinfo {year}
  {2020}),\ \Eprint {https://arxiv.org/abs/2001.02796} {arXiv:2001.02796
  [cond-mat.mes-hall]} \BibitemShut {NoStop}%
\bibitem [{\citenamefont {Xu}\ \emph {et~al.}(2021)\citenamefont {Xu},
  \citenamefont {Ray}, \citenamefont {Shao}, \citenamefont {Jiang},
  \citenamefont {Lee}, \citenamefont {Weber}, \citenamefont {Goldberger},
  \citenamefont {Watanabe}, \citenamefont {Taniguchi}, \citenamefont {Muller},
  \citenamefont {Mak},\ and\ \citenamefont {Shan}}]{Xu2021}%
  \BibitemOpen
  \bibfield  {author} {\bibinfo {author} {\bibfnamefont {Y.}~\bibnamefont
  {Xu}}, \bibinfo {author} {\bibfnamefont {A.}~\bibnamefont {Ray}}, \bibinfo
  {author} {\bibfnamefont {Y.-T.}\ \bibnamefont {Shao}}, \bibinfo {author}
  {\bibfnamefont {S.}~\bibnamefont {Jiang}}, \bibinfo {author} {\bibfnamefont
  {K.}~\bibnamefont {Lee}}, \bibinfo {author} {\bibfnamefont {D.}~\bibnamefont
  {Weber}}, \bibinfo {author} {\bibfnamefont {J.~E.}\ \bibnamefont
  {Goldberger}}, \bibinfo {author} {\bibfnamefont {K.}~\bibnamefont
  {Watanabe}}, \bibinfo {author} {\bibfnamefont {T.}~\bibnamefont {Taniguchi}},
  \bibinfo {author} {\bibfnamefont {D.~A.}\ \bibnamefont {Muller}}, \bibinfo
  {author} {\bibfnamefont {K.~F.}\ \bibnamefont {Mak}},\ and\ \bibinfo {author}
  {\bibfnamefont {J.}~\bibnamefont {Shan}},\ }\href
  {https://doi.org/10.1038/s41565-021-01014-y} {\bibfield  {journal} {\bibinfo
  {journal} {Nature Nanotechnology}\ } (\bibinfo {year} {2021})}\BibitemShut
  {NoStop}%
\bibitem [{\citenamefont {Alexeev}\ \emph {et~al.}(2019)\citenamefont
  {Alexeev}, \citenamefont {Ruiz-Tijerina}, \citenamefont {Danovich},
  \citenamefont {Hamer}, \citenamefont {Terry}, \citenamefont {Nayak},
  \citenamefont {Ahn}, \citenamefont {Pak}, \citenamefont {Lee}, \citenamefont
  {Sohn}, \citenamefont {Molas}, \citenamefont {Koperski}, \citenamefont
  {Watanabe}, \citenamefont {Taniguchi}, \citenamefont {Novoselov},
  \citenamefont {Gorbachev}, \citenamefont {Shin}, \citenamefont {Fal'ko},\
  and\ \citenamefont {Tartakovskii}}]{Alexeev2019}%
  \BibitemOpen
  \bibfield  {author} {\bibinfo {author} {\bibfnamefont {E.~M.}\ \bibnamefont
  {Alexeev}}, \bibinfo {author} {\bibfnamefont {D.~A.}\ \bibnamefont
  {Ruiz-Tijerina}}, \bibinfo {author} {\bibfnamefont {M.}~\bibnamefont
  {Danovich}}, \bibinfo {author} {\bibfnamefont {M.~J.}\ \bibnamefont {Hamer}},
  \bibinfo {author} {\bibfnamefont {D.~J.}\ \bibnamefont {Terry}}, \bibinfo
  {author} {\bibfnamefont {P.~K.}\ \bibnamefont {Nayak}}, \bibinfo {author}
  {\bibfnamefont {S.}~\bibnamefont {Ahn}}, \bibinfo {author} {\bibfnamefont
  {S.}~\bibnamefont {Pak}}, \bibinfo {author} {\bibfnamefont {J.}~\bibnamefont
  {Lee}}, \bibinfo {author} {\bibfnamefont {J.~I.}\ \bibnamefont {Sohn}},
  \bibinfo {author} {\bibfnamefont {M.~R.}\ \bibnamefont {Molas}}, \bibinfo
  {author} {\bibfnamefont {M.}~\bibnamefont {Koperski}}, \bibinfo {author}
  {\bibfnamefont {K.}~\bibnamefont {Watanabe}}, \bibinfo {author}
  {\bibfnamefont {T.}~\bibnamefont {Taniguchi}}, \bibinfo {author}
  {\bibfnamefont {K.~S.}\ \bibnamefont {Novoselov}}, \bibinfo {author}
  {\bibfnamefont {R.~V.}\ \bibnamefont {Gorbachev}}, \bibinfo {author}
  {\bibfnamefont {H.~S.}\ \bibnamefont {Shin}}, \bibinfo {author}
  {\bibfnamefont {V.~I.}\ \bibnamefont {Fal'ko}},\ and\ \bibinfo {author}
  {\bibfnamefont {A.~I.}\ \bibnamefont {Tartakovskii}},\ }\href
  {https://doi.org/10.1038/s41586-019-0986-9} {\bibfield  {journal} {\bibinfo
  {journal} {Nature}\ }\textbf {\bibinfo {volume} {567}},\ \bibinfo {pages}
  {81} (\bibinfo {year} {2019})}\BibitemShut {NoStop}%
\bibitem [{\citenamefont {Glazov}\ \emph {et~al.}(2014)\citenamefont {Glazov},
  \citenamefont {Amand}, \citenamefont {Marie}, \citenamefont {Lagarde},
  \citenamefont {Bouet},\ and\ \citenamefont {Urbaszek}}]{Glazov2014}%
  \BibitemOpen
  \bibfield  {author} {\bibinfo {author} {\bibfnamefont {M.~M.}\ \bibnamefont
  {Glazov}}, \bibinfo {author} {\bibfnamefont {T.}~\bibnamefont {Amand}},
  \bibinfo {author} {\bibfnamefont {X.}~\bibnamefont {Marie}}, \bibinfo
  {author} {\bibfnamefont {D.}~\bibnamefont {Lagarde}}, \bibinfo {author}
  {\bibfnamefont {L.}~\bibnamefont {Bouet}},\ and\ \bibinfo {author}
  {\bibfnamefont {B.}~\bibnamefont {Urbaszek}},\ }\href
  {https://link.aps.org/doi/10.1103/PhysRevB.89.201302} {\bibfield  {journal}
  {\bibinfo  {journal} {Phys. Rev. B}\ }\textbf {\bibinfo {volume} {89}},\
  \bibinfo {pages} {201302(R)} (\bibinfo {year} {2014})}\BibitemShut {NoStop}%
\bibitem [{\citenamefont {Yu}\ \emph {et~al.}(2014)\citenamefont {Yu},
  \citenamefont {Liu}, \citenamefont {Gong}, \citenamefont {Xu},\ and\
  \citenamefont {Yao}}]{WangYao2014}%
  \BibitemOpen
  \bibfield  {author} {\bibinfo {author} {\bibfnamefont {H.}~\bibnamefont
  {Yu}}, \bibinfo {author} {\bibfnamefont {G.-B.}\ \bibnamefont {Liu}},
  \bibinfo {author} {\bibfnamefont {P.}~\bibnamefont {Gong}}, \bibinfo {author}
  {\bibfnamefont {X.}~\bibnamefont {Xu}},\ and\ \bibinfo {author}
  {\bibfnamefont {W.}~\bibnamefont {Yao}},\ }\href
  {https://doi.org/10.1038/ncomms4876} {\bibfield  {journal} {\bibinfo
  {journal} {Nature Communications}\ }\textbf {\bibinfo {volume} {5}},\
  \bibinfo {pages} {3876} (\bibinfo {year} {2014})}\BibitemShut {NoStop}%
\bibitem [{\citenamefont {Qiu}\ \emph {et~al.}(2015)\citenamefont {Qiu},
  \citenamefont {Cao},\ and\ \citenamefont {Louie}}]{Qiu2015}%
  \BibitemOpen
  \bibfield  {author} {\bibinfo {author} {\bibfnamefont {D.~Y.}\ \bibnamefont
  {Qiu}}, \bibinfo {author} {\bibfnamefont {T.}~\bibnamefont {Cao}},\ and\
  \bibinfo {author} {\bibfnamefont {S.~G.}\ \bibnamefont {Louie}},\ }\href
  {https://doi.org/10.1103/PhysRevLett.115.176801} {\bibfield  {journal}
  {\bibinfo  {journal} {Phys. Rev. Lett.}\ }\textbf {\bibinfo {volume} {115}},\
  \bibinfo {pages} {176801} (\bibinfo {year} {2015})}\BibitemShut {NoStop}%
\bibitem [{\citenamefont {Srivastava}\ \emph {et~al.}(2015)\citenamefont
  {Srivastava}, \citenamefont {Sidler}, \citenamefont {Allain}, \citenamefont
  {Lembke}, \citenamefont {Kis},\ and\ \citenamefont
  {Imamoğlu}}]{Srivastava2015}%
  \BibitemOpen
  \bibfield  {author} {\bibinfo {author} {\bibfnamefont {A.}~\bibnamefont
  {Srivastava}}, \bibinfo {author} {\bibfnamefont {M.}~\bibnamefont {Sidler}},
  \bibinfo {author} {\bibfnamefont {A.~V.}\ \bibnamefont {Allain}}, \bibinfo
  {author} {\bibfnamefont {D.~S.}\ \bibnamefont {Lembke}}, \bibinfo {author}
  {\bibfnamefont {A.}~\bibnamefont {Kis}},\ and\ \bibinfo {author}
  {\bibfnamefont {A.}~\bibnamefont {Imamoğlu}},\ }\href
  {http://dx.doi.org/10.1038/nphys3203} {\bibfield  {journal} {\bibinfo
  {journal} {Nature Physics}\ }\textbf {\bibinfo {volume} {11}},\ \bibinfo
  {pages} {141–147} (\bibinfo {year} {2015})}\BibitemShut {NoStop}%
\bibitem [{\citenamefont {Wu}\ \emph {et~al.}(2016)\citenamefont {Wu},
  \citenamefont {Xu}, \citenamefont {Lu}, \citenamefont {Khamoshi},
  \citenamefont {Liu}, \citenamefont {Han}, \citenamefont {Wu}, \citenamefont
  {Lin}, \citenamefont {Long}, \citenamefont {He},\ and\ \citenamefont
  {et~al.}}]{Xu2015}%
  \BibitemOpen
  \bibfield  {author} {\bibinfo {author} {\bibfnamefont {Z.}~\bibnamefont
  {Wu}}, \bibinfo {author} {\bibfnamefont {S.}~\bibnamefont {Xu}}, \bibinfo
  {author} {\bibfnamefont {H.}~\bibnamefont {Lu}}, \bibinfo {author}
  {\bibfnamefont {A.}~\bibnamefont {Khamoshi}}, \bibinfo {author}
  {\bibfnamefont {G.-B.}\ \bibnamefont {Liu}}, \bibinfo {author} {\bibfnamefont
  {T.}~\bibnamefont {Han}}, \bibinfo {author} {\bibfnamefont {Y.}~\bibnamefont
  {Wu}}, \bibinfo {author} {\bibfnamefont {J.}~\bibnamefont {Lin}}, \bibinfo
  {author} {\bibfnamefont {G.}~\bibnamefont {Long}}, \bibinfo {author}
  {\bibfnamefont {Y.}~\bibnamefont {He}},\ and\ \bibinfo {author} {\bibnamefont
  {et~al.}},\ }\bibfield  {journal} {\bibinfo  {journal} {Nature
  Communications}\ }\textbf {\bibinfo {volume} {7}},\ \href
  {https://doi.org/10.1038/ncomms12955} {10.1038/ncomms12955} (\bibinfo {year}
  {2016})\BibitemShut {NoStop}%
\bibitem [{\citenamefont {Li}\ \emph {et~al.}(2014)\citenamefont {Li},
  \citenamefont {Ludwig}, \citenamefont {Low}, \citenamefont {Chernikov},
  \citenamefont {Cui}, \citenamefont {Arefe}, \citenamefont {Kim},
  \citenamefont {van~der Zande}, \citenamefont {Rigosi}, \citenamefont {Hill},
  \citenamefont {Kim}, \citenamefont {Hone}, \citenamefont {Li}, \citenamefont
  {Smirnov},\ and\ \citenamefont {Heinz}}]{Heinz2014}%
  \BibitemOpen
  \bibfield  {author} {\bibinfo {author} {\bibfnamefont {Y.}~\bibnamefont
  {Li}}, \bibinfo {author} {\bibfnamefont {J.}~\bibnamefont {Ludwig}}, \bibinfo
  {author} {\bibfnamefont {T.}~\bibnamefont {Low}}, \bibinfo {author}
  {\bibfnamefont {A.}~\bibnamefont {Chernikov}}, \bibinfo {author}
  {\bibfnamefont {X.}~\bibnamefont {Cui}}, \bibinfo {author} {\bibfnamefont
  {G.}~\bibnamefont {Arefe}}, \bibinfo {author} {\bibfnamefont {Y.~D.}\
  \bibnamefont {Kim}}, \bibinfo {author} {\bibfnamefont {A.~M.}\ \bibnamefont
  {van~der Zande}}, \bibinfo {author} {\bibfnamefont {A.}~\bibnamefont
  {Rigosi}}, \bibinfo {author} {\bibfnamefont {H.~M.}\ \bibnamefont {Hill}},
  \bibinfo {author} {\bibfnamefont {S.~H.}\ \bibnamefont {Kim}}, \bibinfo
  {author} {\bibfnamefont {J.}~\bibnamefont {Hone}}, \bibinfo {author}
  {\bibfnamefont {Z.}~\bibnamefont {Li}}, \bibinfo {author} {\bibfnamefont
  {D.}~\bibnamefont {Smirnov}},\ and\ \bibinfo {author} {\bibfnamefont {T.~F.}\
  \bibnamefont {Heinz}},\ }\href
  {https://doi.org/10.1103/PhysRevLett.113.266804} {\bibfield  {journal}
  {\bibinfo  {journal} {Phys. Rev. Lett.}\ }\textbf {\bibinfo {volume} {113}},\
  \bibinfo {pages} {266804} (\bibinfo {year} {2014})}\BibitemShut {NoStop}%
\bibitem [{\citenamefont {Yu}\ \emph {et~al.}(2015)\citenamefont {Yu},
  \citenamefont {Cui}, \citenamefont {Xu},\ and\ \citenamefont
  {Yao}}]{WangYao_Review}%
  \BibitemOpen
  \bibfield  {author} {\bibinfo {author} {\bibfnamefont {H.}~\bibnamefont
  {Yu}}, \bibinfo {author} {\bibfnamefont {X.}~\bibnamefont {Cui}}, \bibinfo
  {author} {\bibfnamefont {X.}~\bibnamefont {Xu}},\ and\ \bibinfo {author}
  {\bibfnamefont {W.}~\bibnamefont {Yao}},\ }\href
  {https://doi.org/10.1093/nsr/nwu078} {\bibfield  {journal} {\bibinfo
  {journal} {National Science Review}\ }\textbf {\bibinfo {volume} {2}},\
  \bibinfo {pages} {57} (\bibinfo {year} {2015})}\BibitemShut {NoStop}%
\bibitem [{\citenamefont {Smoleński}\ \emph {et~al.}(2021)\citenamefont
  {Smoleński}, \citenamefont {Dolgirev}, \citenamefont {Kuhlenkamp},
  \citenamefont {Popert}, \citenamefont {Shimazaki}, \citenamefont {Back},
  \citenamefont {Lu}, \citenamefont {Kroner}, \citenamefont {Watanabe},
  \citenamefont {Taniguchi},\ and\ \citenamefont {et~al.}}]{Smolenski2021}%
  \BibitemOpen
  \bibfield  {author} {\bibinfo {author} {\bibfnamefont {T.}~\bibnamefont
  {Smoleński}}, \bibinfo {author} {\bibfnamefont {P.~E.}\ \bibnamefont
  {Dolgirev}}, \bibinfo {author} {\bibfnamefont {C.}~\bibnamefont
  {Kuhlenkamp}}, \bibinfo {author} {\bibfnamefont {A.}~\bibnamefont {Popert}},
  \bibinfo {author} {\bibfnamefont {Y.}~\bibnamefont {Shimazaki}}, \bibinfo
  {author} {\bibfnamefont {P.}~\bibnamefont {Back}}, \bibinfo {author}
  {\bibfnamefont {X.}~\bibnamefont {Lu}}, \bibinfo {author} {\bibfnamefont
  {M.}~\bibnamefont {Kroner}}, \bibinfo {author} {\bibfnamefont
  {K.}~\bibnamefont {Watanabe}}, \bibinfo {author} {\bibfnamefont
  {T.}~\bibnamefont {Taniguchi}},\ and\ \bibinfo {author} {\bibnamefont
  {et~al.}},\ }\href {https://doi.org/10.1038/s41586-021-03590-4} {\bibfield
  {journal} {\bibinfo  {journal} {Nature}\ }\textbf {\bibinfo {volume} {595}},\
  \bibinfo {pages} {53–57} (\bibinfo {year} {2021})}\BibitemShut {NoStop}%
\bibitem [{\citenamefont {Maletinsky}\ \emph {et~al.}(2012)\citenamefont
  {Maletinsky}, \citenamefont {Hong}, \citenamefont {Grinolds}, \citenamefont
  {Hausmann}, \citenamefont {Lukin}, \citenamefont {Walsworth}, \citenamefont
  {Loncar},\ and\ \citenamefont {Yacoby}}]{Maletinsky2012}%
  \BibitemOpen
  \bibfield  {author} {\bibinfo {author} {\bibfnamefont {P.}~\bibnamefont
  {Maletinsky}}, \bibinfo {author} {\bibfnamefont {S.}~\bibnamefont {Hong}},
  \bibinfo {author} {\bibfnamefont {M.~S.}\ \bibnamefont {Grinolds}}, \bibinfo
  {author} {\bibfnamefont {B.}~\bibnamefont {Hausmann}}, \bibinfo {author}
  {\bibfnamefont {M.~D.}\ \bibnamefont {Lukin}}, \bibinfo {author}
  {\bibfnamefont {R.~L.}\ \bibnamefont {Walsworth}}, \bibinfo {author}
  {\bibfnamefont {M.}~\bibnamefont {Loncar}},\ and\ \bibinfo {author}
  {\bibfnamefont {A.}~\bibnamefont {Yacoby}},\ }\href
  {https://doi.org/10.1038/nnano.2012.50} {\bibfield  {journal} {\bibinfo
  {journal} {Nature Nanotechnology}\ }\textbf {\bibinfo {volume} {7}},\
  \bibinfo {pages} {320} (\bibinfo {year} {2012})}\BibitemShut {NoStop}%
\bibitem [{Note1()}]{Note1}%
  \BibitemOpen
  \bibinfo {note} {Since the strength of the electron-hole exchange interaction
  depends on the dielectric constant, its precise value will strongly vary for
  each system. For $\protect \mathcal {J}\gtrsim \SI {250}{meV}$, $\lambda
  _{\protect \bm {G},l}\gg \lambda _{\protect \bm {G},t}$ and the longitudinal
  modes couple weakly to the zero-momentum states due to the large energy
  detuning. Consequently, there are only two bright Umklapp exciton resonances
  corresponding to the transverse polarized exciton branch. For $\protect
  \mathcal {J} \lesssim 50$~meV, $\lambda _{\protect \bm {G},l} \sim \lambda
  _{\protect \bm {G},t}$ and the longitudinal modes also become bright,
  resulting in four Umklapp resonances.}\BibitemShut {Stop}%
\bibitem [{\citenamefont {Chen}\ \emph {et~al.}(2020)\citenamefont {Chen},
  \citenamefont {Sharpe}, \citenamefont {Fox}, \citenamefont {Wang},
  \citenamefont {Lyu}, \citenamefont {Jiang}, \citenamefont {Li}, \citenamefont
  {Watanabe}, \citenamefont {Taniguchi}, \citenamefont {Crommie}, \citenamefont
  {Kastner}, \citenamefont {Shi}, \citenamefont {Goldhaber-Gordon},
  \citenamefont {Zhang},\ and\ \citenamefont {Wang}}]{Feng-Wang2020}%
  \BibitemOpen
  \bibfield  {author} {\bibinfo {author} {\bibfnamefont {G.}~\bibnamefont
  {Chen}}, \bibinfo {author} {\bibfnamefont {A.~L.}\ \bibnamefont {Sharpe}},
  \bibinfo {author} {\bibfnamefont {E.~J.}\ \bibnamefont {Fox}}, \bibinfo
  {author} {\bibfnamefont {S.}~\bibnamefont {Wang}}, \bibinfo {author}
  {\bibfnamefont {B.}~\bibnamefont {Lyu}}, \bibinfo {author} {\bibfnamefont
  {L.}~\bibnamefont {Jiang}}, \bibinfo {author} {\bibfnamefont
  {H.}~\bibnamefont {Li}}, \bibinfo {author} {\bibfnamefont {K.}~\bibnamefont
  {Watanabe}}, \bibinfo {author} {\bibfnamefont {T.}~\bibnamefont {Taniguchi}},
  \bibinfo {author} {\bibfnamefont {M.~F.}\ \bibnamefont {Crommie}}, \bibinfo
  {author} {\bibfnamefont {M.~A.}\ \bibnamefont {Kastner}}, \bibinfo {author}
  {\bibfnamefont {Z.}~\bibnamefont {Shi}}, \bibinfo {author} {\bibfnamefont
  {D.}~\bibnamefont {Goldhaber-Gordon}}, \bibinfo {author} {\bibfnamefont
  {Y.}~\bibnamefont {Zhang}},\ and\ \bibinfo {author} {\bibfnamefont
  {F.}~\bibnamefont {Wang}},\ }\href@noop {} {\bibinfo {title} {Tunable
  ferromagnetism at non-integer filling of a moir\'e superlattice}} (\bibinfo
  {year} {2020}),\ \Eprint {https://arxiv.org/abs/2012.10075} {arXiv:2012.10075
  [cond-mat.mes-hall]} \BibitemShut {NoStop}%
\bibitem [{\citenamefont {Daveau}\ \emph {et~al.}(2020)\citenamefont {Daveau},
  \citenamefont {Vandekerckhove}, \citenamefont {Mukherjee}, \citenamefont
  {Wang}, \citenamefont {Shan}, \citenamefont {Mak}, \citenamefont
  {Vamivakas},\ and\ \citenamefont {Fuchs}}]{Mak-Shan2020}%
  \BibitemOpen
  \bibfield  {author} {\bibinfo {author} {\bibfnamefont {R.~S.}\ \bibnamefont
  {Daveau}}, \bibinfo {author} {\bibfnamefont {T.}~\bibnamefont
  {Vandekerckhove}}, \bibinfo {author} {\bibfnamefont {A.}~\bibnamefont
  {Mukherjee}}, \bibinfo {author} {\bibfnamefont {Z.}~\bibnamefont {Wang}},
  \bibinfo {author} {\bibfnamefont {J.}~\bibnamefont {Shan}}, \bibinfo {author}
  {\bibfnamefont {K.~F.}\ \bibnamefont {Mak}}, \bibinfo {author} {\bibfnamefont
  {A.~N.}\ \bibnamefont {Vamivakas}},\ and\ \bibinfo {author} {\bibfnamefont
  {G.~D.}\ \bibnamefont {Fuchs}},\ }\href {https://doi.org/10.1063/5.0013825}
  {\bibfield  {journal} {\bibinfo  {journal} {APL Photonics}\ }\textbf
  {\bibinfo {volume} {5}},\ \bibinfo {pages} {096105} (\bibinfo {year}
  {2020})}\BibitemShut {NoStop}%
\bibitem [{\citenamefont {Zhou}\ \emph {et~al.}(2021)\citenamefont {Zhou},
  \citenamefont {Sung}, \citenamefont {Brutschea}, \citenamefont {Esterlis},
  \citenamefont {Wang}, \citenamefont {Scuri}, \citenamefont {Gelly},
  \citenamefont {Heo}, \citenamefont {Taniguchi}, \citenamefont {Watanabe},
  \citenamefont {Zar{\'a}nd}, \citenamefont {Lukin}, \citenamefont {Kim},
  \citenamefont {Demler},\ and\ \citenamefont {Park}}]{Zhou21}%
  \BibitemOpen
  \bibfield  {author} {\bibinfo {author} {\bibfnamefont {Y.}~\bibnamefont
  {Zhou}}, \bibinfo {author} {\bibfnamefont {J.}~\bibnamefont {Sung}}, \bibinfo
  {author} {\bibfnamefont {E.}~\bibnamefont {Brutschea}}, \bibinfo {author}
  {\bibfnamefont {I.}~\bibnamefont {Esterlis}}, \bibinfo {author}
  {\bibfnamefont {Y.}~\bibnamefont {Wang}}, \bibinfo {author} {\bibfnamefont
  {G.}~\bibnamefont {Scuri}}, \bibinfo {author} {\bibfnamefont {R.~J.}\
  \bibnamefont {Gelly}}, \bibinfo {author} {\bibfnamefont {H.}~\bibnamefont
  {Heo}}, \bibinfo {author} {\bibfnamefont {T.}~\bibnamefont {Taniguchi}},
  \bibinfo {author} {\bibfnamefont {K.}~\bibnamefont {Watanabe}}, \bibinfo
  {author} {\bibfnamefont {G.}~\bibnamefont {Zar{\'a}nd}}, \bibinfo {author}
  {\bibfnamefont {M.~D.}\ \bibnamefont {Lukin}}, \bibinfo {author}
  {\bibfnamefont {P.}~\bibnamefont {Kim}}, \bibinfo {author} {\bibfnamefont
  {E.}~\bibnamefont {Demler}},\ and\ \bibinfo {author} {\bibfnamefont
  {H.}~\bibnamefont {Park}},\ }\href
  {https://doi.org/10.1038/s41586-021-03560-w} {\bibfield  {journal} {\bibinfo
  {journal} {Nature}\ }\textbf {\bibinfo {volume} {595}},\ \bibinfo {pages}
  {48} (\bibinfo {year} {2021})}\BibitemShut {NoStop}%
\bibitem [{\citenamefont {Fey}\ \emph {et~al.}(2020)\citenamefont {Fey},
  \citenamefont {Schmelcher}, \citenamefont {Imamoglu},\ and\ \citenamefont
  {Schmidt}}]{Fey_2020}%
  \BibitemOpen
  \bibfield  {author} {\bibinfo {author} {\bibfnamefont {C.}~\bibnamefont
  {Fey}}, \bibinfo {author} {\bibfnamefont {P.}~\bibnamefont {Schmelcher}},
  \bibinfo {author} {\bibfnamefont {A.}~\bibnamefont {Imamoglu}},\ and\
  \bibinfo {author} {\bibfnamefont {R.}~\bibnamefont {Schmidt}},\ }\href
  {https://doi.org/10.1103/PhysRevB.101.195417} {\bibfield  {journal} {\bibinfo
   {journal} {Phys. Rev. B}\ }\textbf {\bibinfo {volume} {101}},\ \bibinfo
  {pages} {195417} (\bibinfo {year} {2020})}\BibitemShut {NoStop}%
\bibitem [{\citenamefont {Back}\ \emph {et~al.}(2017)\citenamefont {Back},
  \citenamefont {Sidler}, \citenamefont {Cotlet}, \citenamefont {Srivastava},
  \citenamefont {Takemura}, \citenamefont {Kroner},\ and\ \citenamefont
  {Imamo\ifmmode~\breve{g}\else \u{g}\fi{}lu}}]{Back17}%
  \BibitemOpen
  \bibfield  {author} {\bibinfo {author} {\bibfnamefont {P.}~\bibnamefont
  {Back}}, \bibinfo {author} {\bibfnamefont {M.}~\bibnamefont {Sidler}},
  \bibinfo {author} {\bibfnamefont {O.}~\bibnamefont {Cotlet}}, \bibinfo
  {author} {\bibfnamefont {A.}~\bibnamefont {Srivastava}}, \bibinfo {author}
  {\bibfnamefont {N.}~\bibnamefont {Takemura}}, \bibinfo {author}
  {\bibfnamefont {M.}~\bibnamefont {Kroner}},\ and\ \bibinfo {author}
  {\bibfnamefont {A.}~\bibnamefont {Imamo\ifmmode~\breve{g}\else
  \u{g}\fi{}lu}},\ }\href
  {https://link.aps.org/doi/10.1103/PhysRevLett.118.237404} {\bibfield
  {journal} {\bibinfo  {journal} {Phys. Rev. Lett.}\ }\textbf {\bibinfo
  {volume} {118}},\ \bibinfo {pages} {237404} (\bibinfo {year}
  {2017})}\BibitemShut {NoStop}%
\bibitem [{\citenamefont {Morales-Durán}\ \emph {et~al.}(2021)\citenamefont
  {Morales-Durán}, \citenamefont {Hu}, \citenamefont {Potasz},\ and\
  \citenamefont {MacDonald}}]{Morales-Duran21}%
  \BibitemOpen
  \bibfield  {author} {\bibinfo {author} {\bibfnamefont {N.}~\bibnamefont
  {Morales-Durán}}, \bibinfo {author} {\bibfnamefont {N.~C.}\ \bibnamefont
  {Hu}}, \bibinfo {author} {\bibfnamefont {P.}~\bibnamefont {Potasz}},\ and\
  \bibinfo {author} {\bibfnamefont {A.~H.}\ \bibnamefont {MacDonald}},\
  }\href@noop {} {\bibinfo {title} {Non-local interactions in moir\'e hubbard
  systems}} (\bibinfo {year} {2021}),\ \Eprint
  {https://arxiv.org/abs/2108.03313} {arXiv:2108.03313 [cond-mat.str-el]}
  \BibitemShut {NoStop}%
\bibitem [{Sup()}]{Supplement}%
  \BibitemOpen
  \href@noop {} {}\bibinfo {note} {See Supplemental Material at [] for
  additional information on the $C_3$ model derivation, the estimation of the
  intervalley and intravalley exciton-electron coupling constants, brief review
  of moir\'e magnets in twisted bilayers and details on their spectral function
  signatures. The Supplemental Material includes
  \cite{Sivadas_2018}}\BibitemShut {NoStop}%
\bibitem [{\citenamefont {Kuhlenkamp}\ \emph {et~al.}(2021)\citenamefont
  {Kuhlenkamp}, \citenamefont {Knap}, \citenamefont {Wagner}, \citenamefont
  {Schmidt},\ and\ \citenamefont {Imamoglu}}]{Kuhlenkamp21}%
  \BibitemOpen
  \bibfield  {author} {\bibinfo {author} {\bibfnamefont {C.}~\bibnamefont
  {Kuhlenkamp}}, \bibinfo {author} {\bibfnamefont {M.}~\bibnamefont {Knap}},
  \bibinfo {author} {\bibfnamefont {M.}~\bibnamefont {Wagner}}, \bibinfo
  {author} {\bibfnamefont {R.}~\bibnamefont {Schmidt}},\ and\ \bibinfo {author}
  {\bibfnamefont {A.}~\bibnamefont {Imamoglu}},\ }\href
  {https://arxiv.org/abs/2105.01080} {\bibfield  {journal} {\bibinfo  {journal}
  {arXiv preprint arXiv:2105.01080}\ } (\bibinfo {year} {2021})}\BibitemShut
  {NoStop}%
\bibitem [{Note2()}]{Note2}%
  \BibitemOpen
  \bibinfo {note} {In the presence of an external constant magnetic field, the
  main resonances shift due to the Zeeman effect while the Umklapp peaks remain
  almost degenerate}\BibitemShut {NoStop}%
\bibitem [{\citenamefont {Zhong}\ \emph {et~al.}(2020)\citenamefont {Zhong},
  \citenamefont {Seyler}, \citenamefont {Linpeng}, \citenamefont {Wilson},
  \citenamefont {Taniguchi}, \citenamefont {Watanabe}, \citenamefont {McGuire},
  \citenamefont {Fu}, \citenamefont {Xiao}, \citenamefont {Yao},\ and\
  \citenamefont {et~al.}}]{Zhong2020}%
  \BibitemOpen
  \bibfield  {author} {\bibinfo {author} {\bibfnamefont {D.}~\bibnamefont
  {Zhong}}, \bibinfo {author} {\bibfnamefont {K.~L.}\ \bibnamefont {Seyler}},
  \bibinfo {author} {\bibfnamefont {X.}~\bibnamefont {Linpeng}}, \bibinfo
  {author} {\bibfnamefont {N.~P.}\ \bibnamefont {Wilson}}, \bibinfo {author}
  {\bibfnamefont {T.}~\bibnamefont {Taniguchi}}, \bibinfo {author}
  {\bibfnamefont {K.}~\bibnamefont {Watanabe}}, \bibinfo {author}
  {\bibfnamefont {M.~A.}\ \bibnamefont {McGuire}}, \bibinfo {author}
  {\bibfnamefont {K.-M.~C.}\ \bibnamefont {Fu}}, \bibinfo {author}
  {\bibfnamefont {D.}~\bibnamefont {Xiao}}, \bibinfo {author} {\bibfnamefont
  {W.}~\bibnamefont {Yao}},\ and\ \bibinfo {author} {\bibnamefont {et~al.}},\
  }\href {https://doi.org/10.1038/s41565-019-0629-1} {\bibfield  {journal}
  {\bibinfo  {journal} {Nature Nanotechnology}\ }\textbf {\bibinfo {volume}
  {15}},\ \bibinfo {pages} {187–191} (\bibinfo {year} {2020})}\BibitemShut
  {NoStop}%
\bibitem [{\citenamefont {Ciorciaro}\ \emph {et~al.}(2020)\citenamefont
  {Ciorciaro}, \citenamefont {Kroner}, \citenamefont {Watanabe}, \citenamefont
  {Taniguchi},\ and\ \citenamefont {Imamoglu}}]{Livio2020}%
  \BibitemOpen
  \bibfield  {author} {\bibinfo {author} {\bibfnamefont {L.}~\bibnamefont
  {Ciorciaro}}, \bibinfo {author} {\bibfnamefont {M.}~\bibnamefont {Kroner}},
  \bibinfo {author} {\bibfnamefont {K.}~\bibnamefont {Watanabe}}, \bibinfo
  {author} {\bibfnamefont {T.}~\bibnamefont {Taniguchi}},\ and\ \bibinfo
  {author} {\bibfnamefont {A.}~\bibnamefont {Imamoglu}},\ }\href
  {https://doi.org/10.1103/PhysRevLett.124.197401} {\bibfield  {journal}
  {\bibinfo  {journal} {Phys. Rev. Lett.}\ }\textbf {\bibinfo {volume} {124}},\
  \bibinfo {pages} {197401} (\bibinfo {year} {2020})}\BibitemShut {NoStop}%
\bibitem [{\citenamefont {Sivadas}\ \emph {et~al.}(2018)\citenamefont
  {Sivadas}, \citenamefont {Okamoto}, \citenamefont {Xu}, \citenamefont
  {Fennie},\ and\ \citenamefont {Xiao}}]{Sivadas_2018}%
  \BibitemOpen
  \bibfield  {author} {\bibinfo {author} {\bibfnamefont {N.}~\bibnamefont
  {Sivadas}}, \bibinfo {author} {\bibfnamefont {S.}~\bibnamefont {Okamoto}},
  \bibinfo {author} {\bibfnamefont {X.}~\bibnamefont {Xu}}, \bibinfo {author}
  {\bibfnamefont {C.~J.}\ \bibnamefont {Fennie}},\ and\ \bibinfo {author}
  {\bibfnamefont {D.}~\bibnamefont {Xiao}},\ }\href
  {http://dx.doi.org/10.1021/acs.nanolett.8b03321} {\bibfield  {journal}
  {\bibinfo  {journal} {Nano Letters}\ }\textbf {\bibinfo {volume} {18}},\
  \bibinfo {pages} {7658–7664} (\bibinfo {year} {2018})}\BibitemShut
  {NoStop}%
\end{thebibliography}
\end{document}